\documentclass[journal]{IEEEtran}

\ifCLASSINFOpdf
\else
\fi

\usepackage{algorithm}
\usepackage{algorithmic}
\usepackage{amssymb}
\usepackage{amsthm}
\usepackage{bm}
\usepackage{bbm}
\usepackage{booktabs}
\usepackage[switch]{lineno}

\usepackage{amsmath}

\usepackage{graphicx}
\usepackage{amsmath}
\usepackage{booktabs,xcolor,colortbl}
\usepackage{xparse}
\usepackage{float}
\usepackage{subfig}

\newtheorem*{rep@theorem}{\rep@title}
\newcommand{\newreptheorem}[2]{%
\newenvironment{rep#1}[1]{%
 \def\rep@title{#2 \ref{##1}}%
 \begin{rep@theorem}}%
 {\end{rep@theorem}}}
\newreptheorem{theorem}{Theorem}
\newreptheorem{lemma}{Lemma}
\newreptheorem{claim}{Claim}

\usepackage{mathtools}
\usepackage{color}

\newcommand{\zty}[1]{\textcolor{black}{#1}}
\newcommand{\rtwo}[1]{\textcolor{black}{#1}}
\newcommand{\redarrow}{\textcolor{red}{\bm{\uparrow}}}
\newcommand{\greenarrow}{\textcolor{green}{\bm{\downarrow}}}

\usepackage{etoolbox}
\usepackage{multirow}
\usepackage{hyperref}
\hypersetup{
    colorlinks=true,
    linkcolor=blue,
    filecolor=magenta,      
    urlcolor=cyan,
    pdftitle={Overleaf Example},
    pdfpagemode=FullScreen,
    }
\usepackage[noabbrev]{cleveref}

\begin{document}

\title{Transition Propagation Graph Neural Networks for Temporal Networks}

\author{
	Tongya~Zheng,
	Zunlei~Feng,
  Tianli~Zhang,
  Yunzhi~Hao,
	Mingli~Song*,
	Xingen~Wang,
	Xinyu~Wang,
  Ji~Zhao,
	Chun~Chen
	\IEEEcompsocitemizethanks{
	\IEEEcompsocthanksitem Tongya Zheng, Yunzhi Hao, Xingen Wang, and Xinyu Wang are with the College of Computer Science, Zhejiang University, Hangzhou, China. Email: \{tyzheng,ericohyz, newroot, wangxinyu\}@zju.edu.cn\protect
	\IEEEcompsocthanksitem Zunlei Feng, and Tianli Zhang are with the College of Software Technology, Zhejiang University, Hangzhou, China. Email: \{zunleifeng, zhangtianli\}@zju.edu.cn.
	\IEEEcompsocthanksitem Mingli Song is with ZJU-Bangsun Joint Research Center, Zhejiang University, and Shanghai Institute for Advanced Study of Zhejiang University. Email: brooksong@zju.edu.cn\protect
	\protect
	\IEEEcompsocthanksitem Ji Zhao is with Shanghai Pudong Development Bank Co.,Ltd. Shanghai Branch, Shanghai, China. Email: zhaoj02@spdb.com.cn.
	\IEEEcompsocthanksitem Chun Chen is with the College of Computer Science, Zhejiang University, and Peng Cheng Laboratory. Email: chenc@zju.edu.cn\protect
	\IEEEcompsocthanksitem * Corresponding author\protect
  \protect
  \IEEEcompsocthanksitem This manuscript is submitted for the special issue on Deep Neural Networks for Graphs: Theory, Models, Algorithms and Applications of IEEE Transactions on Neural Networks and Learning Systems.
  \protect
  \\
	}
}

\markboth{IEEE TRANSACTIONS ON NEURAL NETWORKS AND LEARNING SYSTEMS, November 2022}%
{Zheng \MakeLowercase{\textit{et al.}}}

\maketitle

\begin{abstract}
  Researchers of temporal networks (e.g., social networks and transaction networks) have been interested in mining dynamic patterns of nodes from their diverse interactions.
  Inspired by recently powerful graph mining methods like skip-gram models and Graph Neural Networks (GNNs), existing approaches focus on generating temporal node embeddings sequentially with nodes' sequential interactions.
  However, the sequential modeling of previous approaches \zty{cannot} handle the transition structure between nodes' neighbors with limited memorization capacity.
  Detailedly, an effective method for the transition structures is required to both model nodes' personalized patterns adaptively and capture node dynamics accordingly.
  In this paper, we propose a method, namely \textbf{T}rans\textbf{I}tion \textbf{P}ropagation \textbf{G}raph \textbf{N}eural \textbf{N}etworks (TIP-GNN), to tackle the challenges of encoding nodes' transition structures.
  The proposed TIP-GNN focuses on the bilevel graph structure in temporal networks: besides the explicit interaction graph, a node's sequential interactions can also be constructed as a transition graph.
  Based on the bilevel graph, TIP-GNN further encodes transition structures by multi-step transition propagation and distills information from neighborhoods by a bilevel graph convolution.
  Experimental results over various temporal networks reveal the efficiency of our TIP-GNN, with at most \rtwo{7.2\%} improvements of accuracy on temporal link prediction.
  Extensive ablation studies further verify the effectiveness and limitations of the transition propagation module.
  Our code is available at \url{https://github.com/doujiang-zheng/TIP-GNN}.
\end{abstract}


\IEEEpeerreviewmaketitle

\section{Introduction}

\IEEEPARstart{T}{emporal} networks are widespread over real-life scenarios: users create, like, and dislike posts in social networks, and employees send, forward, and reply to emails in email networks~\cite{bennett2007netflix,kunegis2013konect,snapnets,network-data}.
Nodes in these temporal networks exhibit their personalized dynamic patterns when keeping interacting with other nodes.
For both researchers and industrial practitioners, it is beneficial to capture node dynamics to predict whether an account is malicious in risk management systems~\cite{kumar2019jodie,wang2020apan}, and recommend interesting items to users in online advertising systems\cite{zuo2018htne,tgat_iclr20}.

\begin{figure}[t]
    \centering
    \includegraphics[width=0.48\textwidth]{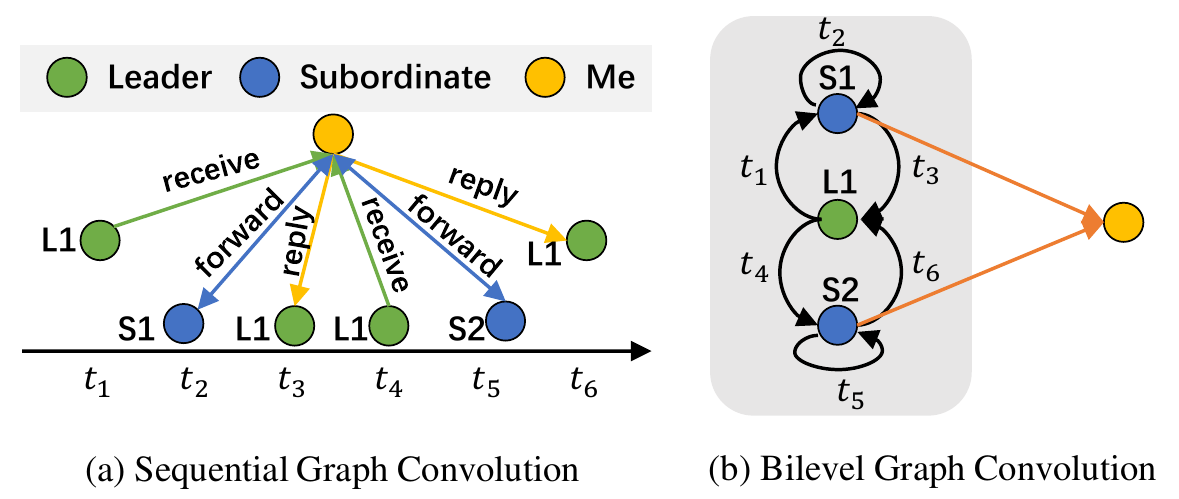}
    \caption{A toy email network where an employee finishes two receive-forward-reply loops with a leader and two subordinates from $t_1$ to $t_6$.
    (a) The sequential graph convolution performs graph convolution sequentially over historical interactions.
    (b) The proposed bilevel graph convolution focuses on the transition structures (black arrows) between neighbors and distills the transition information with a hierarchical attention mechanism (highlighted orange arrows).
    }
    \label{fig:illustration}
\end{figure}

Previous approaches focus on mining node dynamics by aggregating neighborhood features chronologically from nodes' sequential interactions~~\cite{kumar2019jodie,wang2020apan,zuo2018htne,tgat_iclr20,tnode,evolvegcn,zhou2018dynamic,trivedi2018dyrep}, following a generalized sequential graph convolution paradigm as shown in Fig.~\ref{fig:illustration}(a).
These methods are dedicated to modeling neighborhood impacts on nodes while mostly neglecting complex transitions between neighbors, as shown in Fig.~\ref{fig:illustration}(b).
On the one hand, recurrent methods, which can be seen as one-layer temporal Graph Neural Networks (GNNs), update temporal node embeddings iteratively upon new interactions based on specific temporal point processes, such as the triad closure process in DynamicTriad~\cite{zhou2018dynamic}, the attention-based Hawkes process in HTNE~\cite{zuo2018htne}, and the mutual evolution process in JODIE~\cite{kumar2019jodie}. 
On the other hand, temporal GNNs~\cite{wang2020apan,tgat_iclr20,evolvegcn} have been proposed recently to deal with the dynamic high-order structure of the interaction graph, which demonstrate superior performance compared with recurrent methods (shallow GNNs with the one-layer neighborhood).
Overall, existing approaches target generating temporal node embeddings based on the dynamic structure of a temporal network.

However, sequential modeling, the core of existing approaches, can hardly capture the transition structure of nodes' temporal neighbors with limited memorization capacity~\cite{attention2017vaswani}.
Fig.~\ref{fig:illustration} provides an example that an employee finishes two consecutive receive-forward-reply loops in an email network whose loop patterns challenge the representation capability of sequential modeling.
As another example, users in social networks often propagate news in a hierarchical diffusion mechanism.
These nodes' neighbors play different roles in their neighborhoods, presenting complex patterns in chronological order.
If taking the transition structure into account, temporal network methods will make more precise predictions of whether the employee will forward the email to the subordinate or reply to the leader, as shown in Fig.~\ref{fig:illustration}(b).

Nonetheless, there exist two challenges in distilling the transition structure of nodes' neighbors into temporal node embeddings.
Firstly, nodes in a network usually present interactions over a wide range, indicating that their personalized preferences require adaptive modeling.
For instance, a leader may attend several consecutive meetings in one day, while a junior employee may only attend a weekly meeting during workdays.
Thus, temporal node embeddings for the leader and the junior should capture their distinctive structures from their sequential interactions.
Secondly, node dynamics of their temporal interactions imply the variation of their transition structures with time elapsing.
For example, an employee will perform more interactions with high-level colleagues after getting promoted in a company.
Its temporal node embeddings should also evolve forward according to the node dynamics of the transition structure.
An effective method should handle both personalized and dynamic transition structures of neighbors.

To tackle these two challenges mentioned above, we investigate the bilevel graph structure inside temporal networks as shown in Fig.~\ref{fig:illustration}(b).
Besides the explicit interaction graph, a node's sequential interactions can also be constructed as a transition graph according to chronological order.
Specifically, the transition structure of neighbors is then encoded by multi-step propagation via transition links denoted by black arrows in Fig.~\ref{fig:illustration}(b).
The multi-step propagation is proposed to deal with the first challenge of personalized structures: nodes with simple transitions are modeled by short-range propagations, while nodes with complex transitions can benefit from long-range propagations.
Further, the bilevel graph convolution over multi-step neighborhood embeddings (highlighted arrows in Fig.~\ref{fig:illustration}(b)) is proposed to distill node dynamics accordingly from sequential interactions, aiming at the second challenge of dynamic structures.

Concretely, we propose the method, namely \textbf{T}rans\textbf{I}tion \textbf{P}ropagation \textbf{G}raph \textbf{N}eural \textbf{N}etworks (TIP-GNN), to handle nodes' transition structures in temporal networks adaptively and dynamically.
Firstly, TIP-GNN translates a node's interactions into a small transition graph of neighbors and prepares the initialized neighborhood features for propagation.
Secondly, TIP-GNN propagates neighborhood embeddings (latent features) in multiple steps, obtaining both short-range and long-range structures for nodes.
Thirdly, TIP-GNN distills useful information from multi-step neighborhood embeddings by the bilevel graph convolution mechanism: the first transition pooling module accounts for structure encodings of transition structures dynamically, and the second bilevel graph convolution module accounts for detecting transition patterns adaptively.
We conduct experiments over various real-life temporal networks and observe significant improvements in TIP-GNN compared with other state-of-the-art methods on most networks.
Further, we perform extensive ablation studies and parameter sensitivity experiments, validating the effectiveness and limitations of our dedicated transition structures in temporal networks.

Our contributions can be summarized as follows:
\begin{itemize}
  \item Besides the explicit interaction graph of temporal networks, we propose a novel transition graph between nodes' neighbors, revealing the latent dependencies between nodes' neighbors explicitly beyond previous sequential modeling.
  \item To distill transition structures from neighbors, we propose the TIP-GNN, consisting of a transition propagation module and a bilevel graph convolution module, which generates temporal node embeddings adaptively and dynamically with various interactions.
  \item Experimental results demonstrate the robustness and efficiency of our proposed TIP-GNN. Further, we discuss the effectiveness and limitations of the transition structure in temporal networks with a series of ablation studies. 
\end{itemize}

\section{Related Work}

\subsection{Static Graph Embedding}

Learning low-dimensional embeddings has shown great success in many graph mining applications such as risk monitoring~\cite{network2013zhang}, link prediction~\cite{sun2011pathsim}, and item recommendation~\cite{bennett2007netflix} since decades ago~\cite{page1999pagerank,roweis2000nonlinear,belkin2002laplacian,bennett2007netflix,sun2011pathsim}.
Researchers are recently inspired to develop deep graph learning methods~\cite{gnn2020survey,embedding2018survey,survey2020bacciu} with the breakthroughs of deep learning methods in computer vision~\cite{alexnet} and natural language processing~\cite{word2vec}.
The deep graph learning methods mainly consist of two categories: skip-gram models~\cite{embedding2018survey} and graph neural networks~\cite{gnn2020survey}.
DeepWalk~\cite{perozzi2014deepwalk} is the first skip-gram model to model the random walks on graphs as sampled node sequences.
Node2Vec~\cite{grover2016node2vec} further extends DeepWalk with two additional hyper-parameters controlling the preferences of random walks.
Meanwhile, LINE~\cite{tang2015line} explores the high-order similarity of the graph topology between nodes.
On the other hand, graph neural networks~\cite{deffe2016gcn} have made attempts to define the convolution operation in the non-euclidean space.
GCN~\cite{kipf2016semi} simplifies existing GNNs and applies successfully in semi-supervised node classification.
GAT~\cite{velivckovic2017gat} further introduced a self-attention mechanism to compute the weights of neighbors adaptively and achieves substantial improvements in node classification.
Nonetheless, static graph embedding methods are suitable for invariant relationships between entities in real life while being insufficient for temporal networks.

\begin{table}[t]
  \caption{Mathematical symbols and their meanings used in this work.}
  \centering
  \begin{tabular}{l|p{0.8\linewidth}}
    \toprule
    Symbol & Meaning \\ 
    \hline
    \zty{$V, E$} & \zty{the vertex set and edge set of a graph} \\
    \hline
    \zty{$G$} & \zty{$G=(V,E)$, the notation of a graph} \\
    \hline
    $u$ & a node in a graph, $u \in V$ \\
    \hline
    $t$ & a timestamp, $t \in \mathbb{R}^+$ \\
    \hline
    $h_{u,t}^l$ & \zty{$u$'s embedding at $l$-th TIP-GNN layer} \\
    \hline
    \zty{$h_{u}^k$} & \zty{$u$'s embedding at $k$-th propagation step} \\
    \hline
    $S_{u,t}$ & the interaction sequence of node $u$ before $t$ \\
    \hline
    $\mathcal{N}_{u, t}$ & the node set of neighbors in $S_{u,t}$  \\
    \hline
    $s_i$ & an interaction $s_i=(u, v_i, t_i)$ in $S_{u, t}$ \\
    \hline
    $e_i$ & the edge feature of the interaction $s_i$ \\
    \hline
    $A_{u,t}$ & the adjacency matrix of node $u$'s transition graph at $t$ \\
    \hline
    $B_{u, t}$ & the incidence matrix between $\zty{\mathcal{N}_{u,t}}$ and $S_{u,t}$, indicating whether a node $v$ is associated with an interaction $s_i$ \\
    \hline
    $\zty{H_S}$ & \zty{the stacked feature matrix of the interaction set $S_{u,t}$} \\
    \hline
    $\zty{H_{\mathcal{N}_{u,t}}^l}$ & \zty{node embeddings of $\mathcal{N}_{u,t}$ at $l$-th TIP-GNN layer} \\
    \hline
    $v$ & $u$'s neighbor in the set $\mathcal{N}_{u, t}$ \\
    \hline
    $\zty{z_v^k}$ & \zty{node embedding of $u$'s neighbor $v$ at $k$-th propagation step} \\
    \hline
    \zty{$Z_{\mathcal{N}_{u,t}}^k$} & \zty{node embeddings of $\mathcal{N}_{u,t}$ at $k$-th propagation step} \\
    \bottomrule
  \end{tabular}
  \label{tab:symbol}
\end{table}

\subsection{Temporal Network Embedding}

There are numerous temporal networks such as citation and collaboration~\cite{zhou2018dynamic,trivedi2018dyrep,zhang2020tigecmn}, commodity purchasing~\cite{zuo2018htne,kumar2019jodie}, and fraud detection~\cite{wang2020apan}.
According to the summarization of temporal network methods, existing approaches can be divided into two categories based on the property of graphs~\cite{survey2020}.
On the one hand, a discrete-time temporal network refers to a sequence of graph snapshots, where edges in different snapshots may change accordingly.
However, this kind of graph \zty{cannot} model edges at the finest granularity (e.g., at a time scale of seconds)~\cite{nguyen2018ctdne}.
On the other hand, continuous-time graph methods generate the node embeddings dynamically, along with the upcoming new interactions.
HTNE~\cite{zuo2018htne} and DyRep~\cite{trivedi2018dyrep} follow a recurrent paradigm to update node embeddings with historical neighbors.
Additionally, TigeCMN~\cite{zhang2020tigecmn} adopts an external memory network to encode the network dynamics.
Meanwhile, TGAT~\cite{tgat_iclr20} and APAN~\cite{wang2020apan} achieve inductive learning abilities by using graph neural networks.
In comparison, our TIP-GNN proposes a novel transition structure instead of a sequential structure used by previous approaches.

\subsection{Recommender System}

Researches of recommender systems~\cite{timelstm2017zhu,coevolve2016dai,kumar2019jodie,session2019wu,gcsan2019xu,gag2020qiu} are highly related to temporal networks when their interests are modeling user dynamics given historical interactions (e.g., views, clicks, and buys).
However, their two-tower architectures and specific training objectives like BPR loss~\cite{bpr2009rendle} are mainly optimized for recommendation tasks and not suitable for temporal networks that are homogeneous or long time-range.
Methods related to temporal networks can be categorized into two kinds: recurrent recommender systems and session-based recommendations.
On the one hand, Time-LSTM~\cite{timelstm2017zhu}, Deep Coevolve~\cite{coevolve2016dai}, and JODIE~\cite{kumar2019jodie} stand for recurrent recommender systems, ranking items according to nodes' temporal embeddings.
On the other hand, session-based recommendation~\cite{session2019wu,gcsan2019xu,gag2020qiu} explores recommendations based on anonymous user sessions, which utilize an item-to-item graph to encode co-occurrence relationships.
Despite the similarity between the item-to-item graph and our proposed transition graph, session-based recommendation assumes the stationarity on the item-to-item graph, while we use dynamic transition graphs to represent node dynamics.

\begin{figure}[t]
    \centering
    \includegraphics[width=\linewidth]{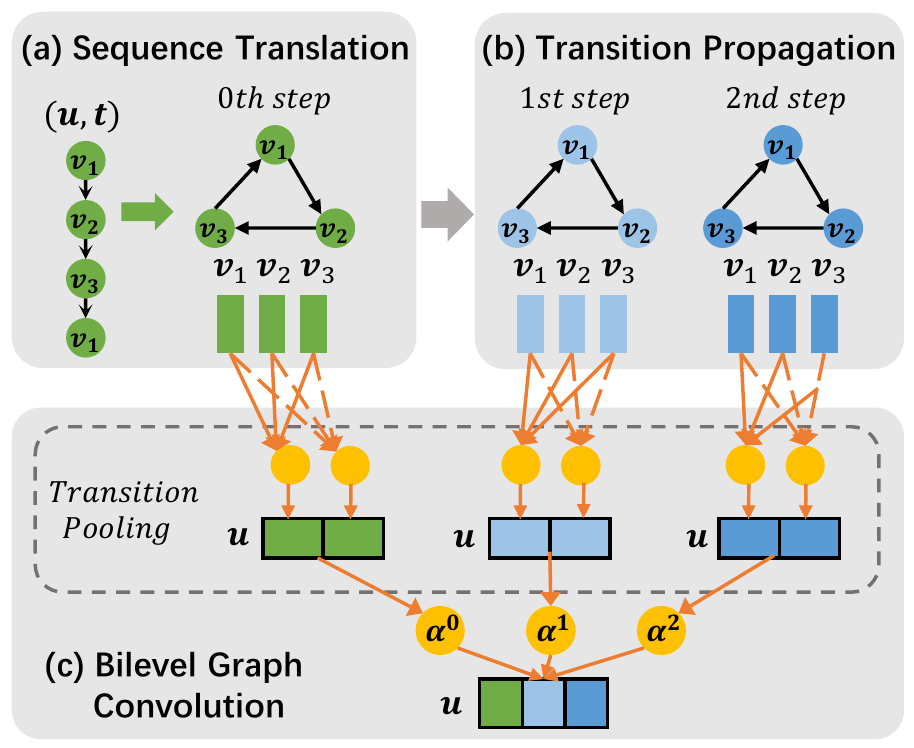}
    \caption{An illustration instance of the one-layer TIP-GNN model.
    (a) Given a query node $\zty{u}$ and timestamp $t$, the sequence translation module translates node $\zty{u}$'s interactions $S_{\zty{u}, t}$ before $t$ into a directed transition graph $A_{\zty{u},t}$, whose initialized node embeddings $\zty{Z_{\mathcal{N}_{u,t}}^0}$ are formed according to Eq.~\ref{eq:init-embed}. 
    (b) The transition propagation module extracts dynamic patterns from the transition graph by propagating node embeddings through multiple steps according to Eq.~\ref{eq:prop-embed}.
    (c) The bilevel graph convolution module distills useful information for node $u$ by a hierarchical attention mechanism: firstly, it performs a multi-head attention pooling at each propagation step according to Eq.~\ref{eq:atte-pooling}; secondly, it aggregates $u$'s node embeddings at different steps by another attention function according to Eq.~\ref{eq:atte-fusion}.
    }
    \label{fig:framework}
\end{figure}

\subsection{Comparisons between TIP-GNN and Previous Works}

TGAT~\cite{tgat_iclr20} and GC-SAN~\cite{gcsan2019xu} are the two most related works to our proposed TIP-GNN.
Firstly, TGAT and TIP-GNN are both based on graph neural networks~\cite{kipf2016semi,hamilton2017graphsage}, and the attention mechanism~\cite{attention2017vaswani}.
However, as depicted in Fig.~\ref{fig:illustration}, TGAT uses sequential graph convolution and can hardly capture the dynamic patterns in the transition graph.
Secondly, GC-SAN and TIP-GNN are both based on the context graph and the attention mechanism~\cite{attention2017vaswani}.
GC-SAN, designed for sequential recommendation, utilizes the item co-occurrence matrix to recommend similar items for users.
In contrast to TIP-GNN, GC-SAN \zty{cannot} mine the high-order relationship in the interaction graph, and \zty{is not} adapted for dynamic patterns as our multi-step transition propagation. 
Specifically, TIP-GNN focuses on nodes' temporal transition graphs and produces temporal node embeddings based on personalized transition graphs.

\section{Method}

Our proposed \textbf{T}rans\textbf{I}tion \textbf{P}ropagation \textbf{G}raph \textbf{N}eural \textbf{N}etworks (TIP-GNN) focuses on the bilevel graph structure of nodes in temporal networks and extracts the transition relationship between nodes' neighbors by bilevel graph neural networks.
As depicted in Fig.~\ref{fig:framework}, TIP-GNN mainly consists of three modules:
the sequence translation module accounts for producing the directed transition graph;
the transition propagation module is designed to represent nodes' dynamic patterns;
the \zty{bilevel graph convolution} module aims at distilling information from the transition graph and the interaction graph.
Finally, our multi-layer TIP-GNN is optimized by the temporal link prediction task.

\subsection{Sequence Translation}

\subsubsection{Transition Graph}

Since proximity structures can be encoded by a graph with an adjacency matrix~\cite{kipf2016semi}, we can also encode transition structures with a transition graph.
Different from a static network, a temporal network denoted by $G=(V,E)$ keeps expanding with time moving forward, where $V$ represents nodes, and $E$ represents edges.
Given a node $u$ and a timestamp $t$, our problem is learning the temporal node embedding $h_{u,t}$, which captures node dynamics precisely from its historical interactions $S_{u,t}$.
We construct the transition graph as a transition matrix $A_{u, t} \in \zty{\vert\mathcal{N}_{u,t}\vert \times \vert\mathcal{N}_{u, t}\vert}$ from these interactions, where $\zty{\mathcal{N}_{u,t} =  \{v_i | (u, v_i, t_i) \in S_{u, t}\}}$ is the neighbor set of node $u$'s interactions.

$S_{u, t} = \{\cdots, s_i, \zty{s_j}, \cdots\}$ denotes the interaction set ordered by timestamps, 
where $s_i$ and $\zty{s_j}$ are two consecutive interactions of the node $u$.
$A_{u, t}$ denotes the transition matrix between node $u$'s neighbors defined by
\begin{equation}
  A_{u,t}[\zty{\text{id}(v_i), \text{id}(v_j)}] = 
  \begin{cases}
    1, & \text{if } \zty{s_i = (u, v_i, t_i), s_j = (u, v_j, t_j)} \\
    0, & \text{otherwise},
  \end{cases}
  \label{eq:trans-adj}
\end{equation}
\zty{where the function $\text{id}(\cdot)$ gives the node index in the neighbor set $\mathcal{N}_{u,t}$.
For instance, Figure~\ref{fig:framework} shows $v_1$ appears at two interactions and $\text{id}(\cdot)$ gives the same node index $v_1$.}
The transition graph $A_{u,t}$ includes only chronologically ascending transition links and excludes bi-directional links because inverse links may obscure the causal relations in the chronological order.  
Moreover, the above definition assumes that $S_{u,t}$ has at least two interactions and makes nodes with sparse interactions less representative in the transition graph.
The self-loop is thus added to the transition graph to help both numerical stability~\cite{kipf2016semi} and interaction sparsity, defined as $\tilde{A}_{u,t} = I + A_{u,t}$.

\subsubsection{Feature Initialization}

To prepare for transition propagation, besides existing node embeddings, we also transform edge features carried by these interactions into node features by an incidence matrix $B_{u, t} \in \zty{\vert\mathcal{N}_{u,t}\vert \times \vert S_{u,t}\vert}$, which connects each interaction $s_i$ with each neighborhood node $\zty{v_i}$.
The incidence matrix $B_{u,t}$ is defined by
\begin{equation}
  B_{u,t}[\zty{\text{id}(v_i)}, i] = 
  \begin{cases}
    1, & \text{if } \zty{s_i = (u, v_i, t_i)} \\
    0, & \text{otherwise},
  \end{cases}
  \label{eq:e2n-incidence}
\end{equation}
where $\zty{\text{id}(v_i)}$ and $i$ are indices of nodes and edges, respectively.
\zty{
  The incidence matrix $B_{u,t}$ could sum up the edge features of neighbor's interactions in order to highlight the numerical impacts of individual neighbors.
}

A specific interaction $\zty{s_i=(u,v_i,t_i)}$ contains its original edge feature $e_i$ and timestamp $t_i$.
Due to the wide range of timespans in different temporal networks (e.g., seconds, days, and weeks), the temporal kernel function of TGAT~\cite{tgat_iclr20} is adopted for robust encoding of timespans, defined by
\begin{equation}
  \Phi(\Delta t) = \textit{concat}(\textit{cos}(\omega_1 \Delta t), \cdots, \textit{cos}(\omega_{d_t} \Delta t)),
  \label{eq:time-encode}
\end{equation}
where $\Delta t = t - t_i$ is the timespan between the prediction timestamp and the edge timestamp, $\omega_1$ is a trainable parameter of the frequency of the cosine function, and $d_t$ is the dimension of the output vector.
The edge features of the interaction $s_i$ is thus defined by $\zty{\tilde{e}_i} = concat(e_i, \Phi(t - t_i))$, and the edge matrix of the interaction set $S_{u,t}$ is stacking interactions' features, defined by $\zty{H_{S} = [\cdots, \tilde{e}_i, \cdots]^\intercal}$.

\zty{
  Let $H_V^l$ be the embeddings of all nodes $V$ in the $l$-th TIP-GNN layer, and then we extract the neighborhood embeddings $H_{\mathcal{N}_{u,t}}^l \in \vert \mathcal{N}_{u,t} \vert \times d$ from the last layer (node features at the 0-th layer), where $d$ is the embedding dimension.
}
Finally, the initialized node embeddings at each layer are calculated by adding the node embeddings and edge features defined by
\begin{equation}
  \zty{
    Z_{\mathcal{N}}^0 = W \times ReLU(W_{n}H_{\mathcal{N}}^{l} + W_e B_{u,t}H_S) + b,
  }
  \label{eq:init-embed}
\end{equation}
where $W, W_n, W_e, b$ are transformation parameters, edges features $\zty{H_S}$ are aggregated through the incidence matrix $B_{u,t}$, 
\zty{$H_{\mathcal{N}}^l$ is short for $H_{\mathcal{N}_{u,t}}^l$}, and $\zty{Z_{\mathcal{N}}^0}$ represents the node embeddings at the 0-th propagation step, as shown in Fig.~\ref{fig:framework}.

\subsection{Transition propagation}

As shown in Fig.~\ref{fig:framework}, the graph propagation module is designed for encoding dynamic patterns from the transition graph.
Inspired by PageRank~\cite{page1999pagerank} and recent graph isomorphism representation works~\cite{gin2018xu}, nodes' dynamic patterns in this paper refer to nodes' complex transitions among nodes' historical neighbors and can be encoded by graph neural networks through multiple propagation steps.
Concretely, the transition propagation rule of a single transition graph is defined by
\begin{equation}
  \zty{
    \tilde{Z}_{\mathcal{N}}^{k+1} = \tilde{A}_{u, t} MLP(Z_{\mathcal{N}}^k),
  }
\end{equation}
where $\zty{k}$ denotes the propagation step, $\zty{Z_{\mathcal{N}}^0}$ is defined in Eq.~\ref{eq:init-embed}, $\tilde{A}_{u,t}$ is the transition matrix with self-loop, and $MLP$ is a multilayer perceptron.
The multilayer perceptron is proposed to enhance the representation capacity of models by its nonlinear transformation.
With $k$ propagation steps, $\zty{Z_{\mathcal{N}}^k}$ can encode $k$-hop subgraphs of the transition graph around neighbors.
Further, a damping factor $\alpha$ is introduced to preserve the uniqueness of node embeddings due to the over-smoothing of graph neural networks, defined by 
\begin{equation}
  \zty{
    Z_{\mathcal{N}}^{k+1} = \alpha Z_{\mathcal{N}}^k + (1 - \alpha) \tilde{Z}_{\mathcal{N}}^{k+1}.
  }
  \label{eq:prop-embed}
\end{equation}
The damping factor~\cite{ppnp2019klicpera} can also be seen as a residual connection between the previous step and the current step, which reduces the learning difficulty of models.

\subsection{Bilevel Graph Convolution}

\subsubsection{Transition Pooling}

The transition pooling module treats neighborhood embeddings at each propagation step separately and produces a summarized graph embedding $\zty{h_{u}^{(k,l+1)}}$ for the specific node $u$.
Specifically, we extract useful information from neighborhood embeddings by performing the attention mechanism~\cite{attention2017vaswani} between node embeddings $h_{u,t}^l$ of the previous \zty{TIP-GNN} layer \zty{(node features at the 0-th layer)} and neighborhood embeddings $\zty{z_v^k \in Z_\mathcal{N}^k}$, defined by 
\begin{equation}
  \begin{split}
    \zty{\alpha_{u v}^{(k,l+1)}} &= \zty{
      \underset{v \in N_{u,t}}{softmax}  \{(W_Q^{l+1} h_{u,t}^l) (W_K^{l+1} z_v^k)^\intercal\},
    } \\
    \zty{h_{u}^{(k,l+1)}} &= \zty{
    \sum_{v \in N_{u,t}} \alpha_{uv}^{(k,l+1)} W_V^{l+1} z_v^k,
    } \\ 
  \end{split}
  \label{eq:atte-pooling}
\end{equation}
where \zty{$h_{u,t}^l$ is extracted from $H_V^l$ (node feature in the first TIP-GNN layer)}, $\zty{W_Q^{l+1}}$ and $\zty{W_K^{l+1}}$ transform node embeddings into the same space, and $\zty{W_V^{l+1}}$ provides additional transformation nonlinearity.
The first equation computes the attention scores between node $u$ and its neighborhood embeddings, and the second equation sums up transformed neighborhood embeddings according to their impacts on the target node $u$.
Moreover, the multi-head attention~\cite{attention2017vaswani} is introduced to enhance the model capacity of graph representation, which computes $\zty{h_u^{(k,l+1)}}$ in Eq.~\ref{eq:atte-pooling} with different parameter matrices several times parallelly.

\subsubsection{Attention Fusion}
The attention fusion module is the second level of the attention mechanism, aiming at fusing node embedding $\zty{h_u^{(k,l+1)}}$ at different propagation steps into a unified representation.
Instead of the similarity-based attention used in Eq.~\ref{eq:atte-pooling}, a projection-based attention is employed here to produce weights for different node embeddings, defined by
\begin{equation}
  \zty{
  \omega_u^{(k,l+1)} = q^\intercal \cdot {sigmoid}(\zty{W^{(k,l+1)} h_u^{(k,l+1)} + b^{(k,l+1)}}), 
  }
\end{equation}
where $\zty{W^{(k,l+1)}}$ and $\zty{b^{(k,l+1)}}$ are specified transformation parameters for each step, $q$ is a shared projection vector across different steps, and $\zty{\omega_u^{(k,l+1)}}$ is an unnormalized scalar weight of each step.
The final node embedding $h_{u,t}^{l+1}$ at $l+1$ \zty{TIP-GNN} layer is computed by
\begin{equation}
  \begin{split}
    \zty{\alpha^{k}_{u}} & \zty{
    = \frac{exp(\omega_u^k)}{\sum_{k'} exp(\omega_u^{k'})},
    } \\
    \zty{h_{u,t}^{l+1}} & \zty{
    = \sum_{k} \alpha_u^{k} h_u^{(k, l+1)}, 
    } \\
  \end{split}
  \label{eq:atte-fusion}
\end{equation}
where $\zty{\alpha_u^k}$ is the normalized weight of each step, and $h_{u,t}^{l+1}$ sums up the weighted node embeddings at different steps.

\subsection{Model Optimization}

\subsubsection{Loss Function}

It is an effective pretext task for generating temporal node embeddings to predict whether a temporal interaction will happen between node pairs in the given timestamp~\cite{wang2020apan,tgat_iclr20}.
The probability $\hat{p}_{uv}^t$ of the interaction's existence between node $u$ and node $v$ at $t$ is thus defined as
\begin{equation}
  \hat{p}_{uv}^t = sigmoid(W\times ReLU(W_u h_{u,t}^L + W_v h_{v,t}^L) + b),
  \label{eq:pred}
\end{equation}
where $W, W_u, W_v$ are transformation matrices, $h_{u,t}^L$ and $h_{v,t}^L$ denote node embeddings at the $L$-th \zty{TIP-GNN} layer, and $L$ is the last \zty{TIP-GNN} layer.
The cross-entropy loss is adopted to classify the existence of the edge, which is defined as follows,
\begin{equation}
  \underset{\theta}{\arg \min} - \sum_{(u, v, t) \in E} \{ log(\hat{p}_{uv}^t) + c \cdot \mathbb{E}_{j \sim P_n(u)} log(1 - \hat{p}_{uj}^t) \},
  \label{eq:loss-ce}
\end{equation}
where $E$ is the set of temporal edges, $P_n(u)$ is a uniform distribution over nodes for negative sampling, and $c$ is the number of negative samples. 
The negative sampling loss is useful for boosting the training efficiency in network embedding methods~\cite{word2vec}.
For simplicity, the number of negative samples is set as 1 in our experiments.
The model parameters are then updated using the Adam~\cite{adam} optimizer, which uses the weight-decay strategy to regularize the magnitude of model parameters.

\subsubsection{Model Size}

The proposed TIP-GNN model introduces additional modules on the transition graph compared with previous methods based on graph neural networks.
Let $d$, $d_e$, $d_t$ be the embedding dimension of node feature, edge feature, and time feature, respectively.
Also, the hidden dimension is all set as $d$ for simplicity.
The first sequential translation module \zty{does not} include trainable parameters.
Let $q$ be the MLP layer in the transition propagation, $K$ be the number of propagation steps, and $L$ be the number of \zty{TIP-GNN} layers.
The parameters of the transition propagation module are $\mathcal{O}(d(d + d_e + d_t)L + qd^2KL)$, which is mainly attributed to the feature initialization and the multilayer perceptron.
The parameters of the bilevel graph convolution module are $\mathcal{O}(3d^2KL + d^2KL + dL)$, which consist of the $K$-step transition pooling and the attention fusion, respectively. 
The final prediction layer also contains $\mathcal{O}(3d^2)$ parameters.
To summarize, the model size of TIP-GNN is $\mathcal{O}(3d^2L + qd^2KL + 4d^2KL + dL + 3d^2)$, where we assume $d, d_e, d_t$ are at the same order.

\subsubsection{Complexity Analysis}

\zty{
  For each node $u$ at $t$, a one-layer TIP-GNN samples $b$ latest interactions from its interaction set $S_{u, t}$, and provides node features, edge features, and edge timestamps. 
  Let $\hat{\mathcal{N}}_{u,t}$ be the sampled neighbor set, a $2$-layer TIP-GNN will sample the second-order interactions recursively from $S_{\hat{\mathcal{N}}_{u,t}}$ following Section 3.4 of TGAT~\cite{tgat_iclr20}.}
\zty{Let the TIP-GNN layer be $L$,} the total number of interactions for a node is denoted by $\mathcal{B}$, which is $1 + b + \cdots + b^L$.
The time complexity of sequence translation is $\mathcal{O}(\mathcal{B})$.
In practice, these transition graphs are cached to boost the training speed.
The time complexity of transition propagation is $\mathcal{O}(3d^2L\mathcal{B} + qd^2KL\mathcal{B} + dKLb\mathcal{B})$, where the last term refers to the propagation step.
The time complexity of the bilevel graph convolution and the prediction layer is $\mathcal{O}(3KLb\mathcal{B} + d^2KL\mathcal{B} + dL\mathcal{B})$ and $\mathcal{O}(3bd^2)$, respectively.
Let $\mathcal{M}$ be the model size of TIP-GNN; the time complexity of TIP-GNN is mainly $\mathcal{O}(\mathcal{B} + b\mathcal{B}\mathcal{M})$, which is $b\mathcal{B}$ times of the model size besides the sequence translation module.

\section{Experiment}
\label{sec:experiment}
We conduct elaborate experiments on various temporal networks to evaluate the efficiency and robustness of TIP-GNN, especially on the effectiveness of the transition propagation module.
We aim to answer the following key research questions:
\begin{itemize}
  \item \textbf{RQ1}: How does TIP-GNN perform compared with other state-of-the-art temporal network methods?
  \item \textbf{RQ2}: What are the effects of our specific transition propagation module on task performance?
  \item \textbf{RQ3}: How do different parameter settings (e.g., the number of \zty{TIP-GNN} layers, the number of attention heads, and so on) affect the TIP-GNN model?
\end{itemize}

\begin{table}[t]
  \caption{Temporal network datasets and statistics. $\vert V \vert$ is the number of nodes in the dataset. $\vert E \vert$ is the number of interactions. The graph density is the ratio between $\vert E \vert$ and $\frac{\vert V \vert (\vert V \vert - 1)}{2}$. The repetition of interactions describes that a node interacts with the same neighbor the last time. The time unit of the timespan is one day.}
  \centering
  \resizebox{1.0\linewidth}{!}{
  \begin{tabular}{lrrlrr}
    \toprule
    Temporal Graph             & $\vert V \vert$ & $\vert E \vert$ & Density  & Repetition & Timespan \\
    \midrule
    \multicolumn{6}{c}{Temporal Link Prediction}   \\
    \midrule
    ia-workplace-contacts      & 92              & 9.8K            & 2.34  & 77.1\%    & 11.43           \\
    ia-contacts-hypertext2009    & 113             & 20.8K           & 3.28  & 59.0\%     & 2.46            \\
    ia-contact  & 274             & 28.2K           & 0.75   & 6.9\%       & 3.97            \\
    fb-forum     & 899             & 33.7K           & 0.08  & 20.8\%    & 164.49          \\
    soc-sign-bitcoin  & 5.8K            & 35.5K           & 0.002   & 0.0\%    & 1903.27         \\
    ia-radoslaw-email      & 167             & 82.9K           & 5.98   & 18.8\%   & 271.19          \\
    ia-movielens-user2tags-10m    & 17K             & 95.5K           & 0.0007 & 19.9\%    & 1108.97         \\
    ia-primary-school-proximity     & 242             & 125.7K          & 4.31  & 38.3\%   & 1.35            \\
    ia-slashdot-reply-dir    & 51K             & 140.7K          & 0.0001 & 4.2\%    & 977.36          \\
    \midrule
    \multicolumn{6}{c}{\zty{Inductive Temporal Link Prediction \& Temporal Node Classification}} \\
    \midrule
    Wikipedia & 9.2K   & 157.4K      & 0.0036  	& 79.1\%   &  29.77        \\
    Reddit			&  10.9K &    672.4K      & 0.011  & 61.4\%  &   31.00    \\
    \bottomrule
  \end{tabular}
  }
  \label{tab:data}
\end{table}

\begin{table*}[htbp]
  \caption{Performance of temporal link prediction. 
  The symbol --- is used for failed experiments.
  Bold font indicates the best performance.
  '*' denotes that TIP-GNN outperforms the best baseline performance statistically significant with $p < 0.05$ under the two-sided t-test.
  The bottom row reports the percentage improvements against the best performance of baseline methods.
  }
  \centering
  \resizebox{1.0\textwidth}{!}{
  \begin{tabular}{lcccccc}
    \toprule
    & \multicolumn{2}{c}{ \textbf{ ia-workplace }} & \multicolumn{2}{c}{ \textbf{ ia-hypertext }} & \multicolumn{2}{c}{ \textbf{ ia-contact }} \\
    \cmidrule(){2-7}
		& Accuracy & AUC & Accuracy & AUC & Accuracy & AUC \\
		\midrule
    Node2Vec~\cite{grover2016node2vec} & $ 0.649 \pm 0.033 $ & $ 0.688 \pm 0.019 $ & $ 0.642 \pm 0.047 $ & $ 0.678 \pm 0.029 $ & $ 0.760 \pm 0.028 $ & $ 0.801 \pm 0.012 $\\
    SAGE~\cite{hamilton2017graphsage} & $ 0.746 \pm 0.008 $ & $ 0.857 \pm 0.008 $ & $ 0.709 \pm 0.009 $ & $ 0.830 \pm 0.006 $ & $ 0.818 \pm 0.004 $ & $ 0.856 \pm 0.003 $\\
    CTDNE~\cite{nguyen2018ctdne} & $ 0.625 \pm 0.014 $ & $ 0.673 \pm 0.010 $ & $ 0.546 \pm 0.069 $ & $ 0.572 \pm 0.036 $ & $ 0.821 \pm 0.006 $ & $ 0.851 \pm 0.004 $\\
    HTNE~\cite{zuo2018htne} & $ 0.621 \pm 0.015 $ & $ 0.661 \pm 0.015 $ & $ 0.517 \pm 0.044 $ & $ 0.540 \pm 0.021 $ & $ 0.806 \pm 0.010 $ & $ 0.831 \pm 0.010 $\\
    TNODE~\cite{tnode} & $ 0.801 \pm 0.009 $ & $ 0.883 \pm 0.006 $ & $ 0.623 \pm 0.071 $ & $ 0.683 \pm 0.019 $ & $ 0.815 \pm 0.008 $ & $ 0.852 \pm 0.002 $\\
    JODIE~\cite{kumar2019jodie} & $ 0.538 \pm 0.011 $ & $ 0.600 \pm 0.054 $ & $ 0.610 \pm 0.035 $ & $ 0.667 \pm 0.067 $ & $ 0.812 \pm 0.005 $ & $ 0.850 \pm 0.005 $\\
    APAN~\cite{wang2020apan} & $ \rtwo{0.695 \pm 0.016} $ & $ \rtwo{0.763 \pm 0.035} $ & $ \rtwo{0.725 \pm 0.007} $ & $ \rtwo{0.807 \pm 0.009} $ & $ \rtwo{0.832 \pm 0.003} $ & $ \rtwo{0.884 \pm 0.002} $\\
    TGAT~\cite{tgat_iclr20} & $ 0.878 \pm 0.001 $ & $ 0.959 \pm 0.001 $ & $ \bm{0.894 \pm 0.001} $ & $ \bm{0.959 \pm 0.001} $ & $ 0.883 \pm 0.001 $ & $ 0.921 \pm 0.001 $\\
    TGN~\cite{tgn2020rossi} & $ 0.883 \pm 0.004 $ & $ 0.960 \pm 0.001 $ & $ \rtwo{0.885 \pm 0.001} $ & $ \rtwo{0.953 \pm 0.001} $ & $ \bm{\rtwo{0.891 \pm 0.001}} $ & $ \bm{\rtwo{0.927 \pm 0.001}} $\\
    TIP-GNN & $ \bm{0.885 \pm 0.010} $ & $  \bm{0.961 \pm 0.007} $ & $  0.889 \pm 0.009 $ & $  0.958 \pm 0.005 $ & $  0.887 \pm 0.003 $ & $ 0.922 \pm 0.003 $\\
    \midrule
    Improvements & {0.2\%} & {0.1\%} & -0.5\% & -0.1\% & {-0.4\%} & {-0.5\%} \\
    \toprule
    & \multicolumn{2}{c}{\textbf{fb-forum}} & \multicolumn{2}{c}{\textbf{soc-bitcoin}} & \multicolumn{2}{c}{\textbf{ia-radoslaw}} \\  
    \cmidrule(){2-7}
		& Accuracy & AUC & Accuracy & AUC & Accuracy & AUC \\
		\midrule
    Node2Vec~\cite{grover2016node2vec} & $ 0.744 \pm 0.009 $ & $ 0.823 \pm 0.008 $ & $ 0.708 \pm 0.014 $ & $ 0.774 \pm 0.012 $ & $ 0.708 \pm 0.011 $ & $ 0.773 \pm 0.010 $\\
    SAGE~\cite{hamilton2017graphsage} & $ 0.636 \pm 0.006 $ & $ 0.724 \pm 0.010 $ & $ 0.651 \pm 0.008 $ & $ 0.734 \pm 0.009 $ & $ 0.804 \pm 0.006 $ & $ 0.894 \pm 0.005 $\\
    CTDNE~\cite{nguyen2018ctdne} & $ 0.745 \pm 0.007 $ & $ 0.817 \pm 0.005 $ & $ 0.778 \pm 0.004 $ & $ 0.836 \pm 0.003 $ & $ 0.723 \pm 0.005 $ & $ 0.795 \pm 0.004 $\\
    HTNE~\cite{zuo2018htne} & $ 0.668 \pm 0.004 $ & $ 0.715 \pm 0.005 $ & $ 0.611 \pm 0.009 $ & $ 0.639 \pm 0.005 $ & $ 0.679 \pm 0.005 $ & $ 0.744 \pm 0.007 $\\
    TNODE~\cite{tnode} & $ 0.716 \pm 0.004 $ & $ 0.795 \pm 0.007 $ & $ 0.724 \pm 0.007 $ & $ 0.793 \pm 0.008 $ & $ 0.775 \pm 0.003 $ & $ 0.863 \pm 0.001 $\\
    JODIE~\cite{kumar2019jodie} & $ 0.632 \pm 0.034 $ & $ 0.751 \pm 0.045 $ & $ 0.814 \pm 0.056 $ & $ 0.880 \pm 0.019 $ & $ 0.713 \pm 0.012 $ & $ 0.784 \pm 0.022 $\\
    APAN~\cite{wang2020apan} & $ \rtwo{0.776 \pm 0.016} $ & $ \rtwo{0.858 \pm 0.018} $ & $ \rtwo{0.775 \pm 0.027} $ & $ \rtwo{0.854 \pm 0.028} $ & $ \rtwo{0.717 \pm 0.018} $ & $ \rtwo{0.794 \pm 0.011} $\\
    TGAT~\cite{tgat_iclr20} & $ 0.794 \pm 0.001 $ & $ 0.877 \pm 0.001 $ & $ \rtwo{0.823 \pm 0.001} $ & $ \rtwo{0.896 \pm 0.001} $ & $ \rtwo{0.841 \pm 0.001} $ & $ \rtwo{0.919 \pm 0.001} $\\
    TGN~\cite{tgn2020rossi} & $ \rtwo{0.783 \pm 0.025} $ & $ \rtwo{0.889 \pm 0.016} $ & $ 0.762 \pm 0.005 $ & $ 0.838 \pm 0.004 $ & $ \rtwo{0.792 \pm 0.012} $ & $ \rtwo{0.885 \pm 0.008} $\\
    TIP-GNN & $ \bm{ 0.824 \pm 0.008 }* $ & $ \bm{ 0.911 \pm 0.006 }* $ & $ \bm{ 0.831 \pm 0.009 } $ & $ \bm{ 0.912 \pm 0.006 }* $ & $ \bm{ 0.857 \pm 0.007 }* $ & $ \bm{ 0.929 \pm 0.005 }* $\\
    \midrule
    Improvements & 3.8\% & \rtwo{2.5\%} & \rtwo{1.0\%} & \rtwo{1.8\%} & \rtwo{1.9\%} & \rtwo{1.1\%} \\
    \toprule
    & \multicolumn{2}{c}{\textbf{ia-movielens}} & \multicolumn{2}{c}{\textbf{ia-primary}} & \multicolumn{2}{c}{\textbf{ia-slashdot}}\\  
		\cmidrule(){2-7}
		& Accuracy & AUC & Accuracy & AUC & Accuracy & AUC \\
		\midrule
    Node2Vec~\cite{grover2016node2vec} & $ 0.696 \pm 0.003 $ & $ 0.756 \pm 0.003 $ & $ 0.586 \pm 0.009 $ & $ 0.622 \pm 0.008 $ & $ 0.723 \pm 0.003 $ & $ 0.800 \pm 0.003 $\\
    SAGE~\cite{hamilton2017graphsage} & $ 0.710 \pm 0.012 $ & $ 0.800 \pm 0.004 $ & $ 0.856 \pm 0.003 $ & $ 0.929 \pm 0.001 $ & $ 0.644 \pm 0.017 $ & $ 0.782 \pm 0.005 $\\
    CTDNE~\cite{nguyen2018ctdne} & $ 0.728 \pm 0.004 $ & $ 0.796 \pm 0.006 $ & $ 0.593 \pm 0.013 $ & $ 0.637 \pm 0.004 $ & $ 0.782 \pm 0.002 $ & $ 0.851 \pm 0.002 $\\
    HTNE~\cite{zuo2018htne} & $ 0.694 \pm 0.005 $ & $ 0.736 \pm 0.003 $ & $ 0.559 \pm 0.008 $ & $ 0.613 \pm 0.004 $ & $ 0.660 \pm 0.003 $ & $ 0.691 \pm 0.002 $\\
    TNODE~\cite{tnode} & $ 0.713 \pm 0.012 $ & $ 0.793 \pm 0.013 $ & $ 0.891 \pm 0.003 $ & $ 0.936 \pm 0.001 $ & $ 0.731 \pm 0.016 $ & $ 0.804 \pm 0.008 $\\
    JODIE~\cite{kumar2019jodie} & $ 0.775 \pm 0.004 $ & $ 0.862 \pm 0.006 $ & $ 0.560 \pm 0.009 $ & $ 0.588 \pm 0.012 $ & $ - $ & $ - $\\
    APAN~\cite{wang2020apan} & $ 0.750 \pm 0.031 $ & $ 0.816 \pm 0.033 $ & $ \rtwo{0.830 \pm 0.048} $ & $ \rtwo{0.911 \pm 0.019} $ & $ \rtwo{0.815 \pm 0.006} $ & $ \rtwo{0.895 \pm 0.003} $\\
    TGAT~\cite{tgat_iclr20} & $ 0.820 \pm 0.001 $ & $ 0.884 \pm 0.001 $ & $ \rtwo{0.921 \pm 0.001} $ & $ \rtwo{0.951 \pm 0.001} $ & $ \rtwo{0.676 \pm 0.001} $ & $ \rtwo{0.805 \pm 0.001} $\\
    TGN~\cite{tgn2020rossi} & $ 0.841 \pm 0.004 $ & $ 0.921 \pm 0.004 $ & $ \rtwo{0.912 \pm 0.021} $ & $ \rtwo{0.961 \pm 0.011} $ & $ 0.789 \pm 0.047 $ & $ 0.896 \pm 0.005 $\\
    TIP-GNN & $ \bm{ 0.847 \pm 0.004 } $ & $ \bm{ 0.922 \pm 0.002 } $ & $ \bm{ 0.939 \pm 0.003 }* $ & $ \bm{ 0.978 \pm 0.001 }* $ & $ \bm{ 0.874 \pm 0.005 }* $ & $ \bm{ 0.938 \pm 0.004 }* $\\
    \midrule
    Improvements & 0.7\% & 0.1\% & \rtwo{2.0\%} & \rtwo{1.8\%} & \rtwo{7.2\%} & \rtwo{4.7\%} \\
    \bottomrule
  \end{tabular}
  }
  \label{tab:auc}
\end{table*}
\begin{table}[t]
    \caption{\zty{Average Precision for temporal link prediction in transductive and inductive settings.}}
    \centering
    \resizebox{\linewidth}{!}{
    \begin{tabular}{lcccc}
        \toprule
        & \multicolumn{2}{c}{Wikipedia} & \multicolumn{2}{c}{Reddit} \\
        \cmidrule{2-5}
        & Trnasductive & Inductive & Transductive & Inductive \\
        \midrule
        Node2Vec & $0.915 \pm 0.003$ & $\dagger$ & $0.846 \pm 0.005$ & $\dagger$ \\
        SAGE & $0.936 \pm 0.003$ & $0.911 \pm 0.003$ & $0.977 \pm 0.002$ & $0.963 \pm 0.002$ \\
        CTDNE & $0.922 \pm 0.005$ & $\dagger$ & $0.914 \pm 0.003$ & $\dagger$ \\
        JODIE & $0.946 \pm 0.005$ & $0.931 \pm 0.004$ & $0.971 \pm 0.003$ & $0.944 \pm 0.011$ \\
        APAN  & $0.981 \pm 0.002$ & $\dagger$ & $\bm{0.992 \pm 0.002}$ & $\dagger$ \\
        TGAT & $0.953 \pm 0.001$ & $0.940 \pm 0.003$ & $0.981 \pm 0.002$ & $0.966 \pm 0.002$ \\
        TGN & $0.985 \pm 0.001$ & $0.978 \pm 0.001$ & $0.987 \pm 0.001$ & $0.976 \pm 0.001$ \\
        TIP-GNN &  $\bm{0.986 \pm 0.001}$ & $\bm{0.982 \pm 0.001}$ & $0.988 \pm 0.001$ & $\bm{0.977 \pm 0.001}$ \\ 
        \bottomrule
    \end{tabular}
    }
    \label{tab:inductive}
\end{table}

\begin{table}[t]
    \caption{AUC scores for temporal node classification. 
    The \textbf{Bold} font represents the best performance, 
    the \underline{underline} font represents the second-best performance, 
    and '*' denotes the statistical significance with $p < 0.05$ under two-sided t-test.}
    \centering
    \resizebox{\linewidth}{!}{
    \begin{tabular}{lcc}
        \toprule
        Methods & Wikipedia & Reddit \\
        \midrule
        Node2Vec & $0.812 \pm 0.018$ & $0.618 \pm 0.050$ \\
        SAGE & $0.824 \pm 0.007$ & $0.612 \pm 0.006$ \\
        CTDNE & $0.759 \pm 0.005$ & $0.594 \pm 0.006$ \\
        HTNE & $0.742 \pm 0.011$ & $0.612 \pm 0.009$ \\
        JODIE & $0.832 \pm 0.005$ & $0.599 \pm 0.021$ \\
        APAN  & $\bm{0.899 \pm 0.003}$* & $0.653 \pm 0.004$ \\
        TGAT & $0.837 \pm 0.007$ & $0.656 \pm 0.007$ \\
        TGN & $0.878 \pm 0.002$ & \underline{$0.671 \pm 0.009$} \\
        \zty{TIP-GNN} & \zty{\underline{$0.880 \pm 0.002$}} & \zty{$\bm{0.697 \pm 0.010}$*} \\
        \bottomrule
    \end{tabular}
    }
    \label{tab:nc}
\end{table}

\subsection{Datasets and Tasks}

Table~\ref{tab:data} lists temporal networks collected over a wide range from Network Repository~\cite{network-data} and SNAP~\cite{kumar2019jodie}.
In consideration of scalability, nine medium-scale networks \zty{without node or edge features} from Network Repository~\cite{network-data} are used for temporal link prediction, which contains various temporal patterns due to huge variations of the timespan.
Meanwhile, two networks with preprocessed edge features from SNAP~\cite{kumar2019jodie} are used for \zty{temporal link prediction in inductive settings and} temporal node classification, whose label ratios are highly imbalanced.
For clarity of analyses, experimental networks are divided according to their node number and graph density.
The network with more than 5,000 nodes is referred to as a \textbf{large} network, and the other networks are seen as \textbf{small} networks.
Meanwhile, the density of a \textbf{sparse} network is less than 0.005, and the other networks are considered as \textbf{dense} networks.

\subsubsection{Temporal Link Prediction}

Temporal networks for temporal link prediction only consist of temporal interactions and do not carry any node features or edge features.
These networks could sufficiently validate the robustness of existing methods because of huge differences in their essential properties.
According to the above divisions of networks, existing networks mainly consist of two kinds, large and sparse networks, and small and dense networks.
Notably, large and sparse networks often spread across a long timespan (over one year), and small and dense networks usually occur during a small timespan (less than two weeks).
Further, networks with mixed properties challenge the representation ability of compared methods.
For example, the timespans of small and dense networks like \emph{fb-forum} and \emph{ia-radoslaw-email} are around half a year.
Experiments on these complex networks could reveal the performance of methods in industrial scenarios to some degree.

Temporal link prediction refers to predicting whether a given node pair will have an interaction at a given timestamp.
We split the datasets into the training, validation, and testing sets with ratios 70:15:15 chronologically.
We use the observed interactions in the testing set as positive samples and obtain negative samples by replacing target nodes in the testing set with nodes that have never interacted with source nodes. 
The labeled datasets are used for the evaluation of temporal network methods.
\zty{
  Our collected temporal graphs from Network Repository~\cite{network-data} provide only graph topology without node or edge features, causing many comparing methods not suitable for new nodes in the testing set.
  Therefore, we remove new nodes from the validation and testing set for a fair comparison.
}

\subsubsection{\zty{Inductive Temporal Link Prediction and Temporal Node Classification}}

Temporal networks for \zty{inductive temporal link prediction and} temporal node classification are two bipartite graphs consisting of user behaviors among Wikipedia pages and subreddits.
Wikipedia consists of 8,227 users and 1,000 most edited pages, while Reddit contains 10,000 active users and 1,000 active subreddits.
The edge features in both Wikipedia and Reddit are text features of interactions, by converting Wikipedia pages and Reddit posts into 172-dimensional vectors under the \emph{linguistic inquiry and word count} (LIWC) categories~\cite{liwc2001}.
Detailedly, the interaction of Wikipedia is that a user-edited a page, while the interaction of Reddit is that a user created a post in the subreddit.

\zty{
  Inductive temporal link prediction validates the inductive capability of TIP-GNN on new nodes in the testing set by hiding 10\% of nodes from the training set following the task settings of TGAT~\cite{tgat_iclr20} and TGN~\cite{tgn2020rossi}.}
Temporal node classification refers to classifying a user's temporal state, whether the user is banned from editing a Wikipedia page or posting in the subreddit.
Also, we split the datasets into the training, validation, and testing sets with ratios 70:15:15 chronologically.
The labels of the additional two datasets are extremely imbalanced: {Wikipedia} has 217 positive labels with 157,474 interactions (=0.14\%), while {Reddit} has 366 true labels among 672,447 interactions (=0.05\%). 

\subsubsection{Evaluation Metrics}
For temporal link prediction, we compute the classification \emph{accuracy} and the \emph{area under the ROC curve} (AUC-ROC) when obtaining the prediction probabilities of edge existence.
\zty{
  For temporal link prediction in transductive and inductive settings, we adopt the Average Precision (AP) following TGAT~\cite{tgat_iclr20} and TGN~\cite{tgn2020rossi}.
}
For temporal node classification, we use the AUC-ROC due to the highly imbalanced labels.

\subsection{Experiment Setting}

\subsubsection{Baseline Methods} 

Most baseline methods generate temporal node embeddings with the pretext task of temporal link prediction.
We employ a logistic classifier for temporal link prediction if methods \zty{cannot} predict temporal links directly.
These methods can be categorized as follows:
\begin{itemize}
  \item \textbf{Static graph methods.} Node2Vec~\cite{grover2016node2vec} and SAGE~\cite{hamilton2017graphsage} are two classic and effective algorithms for skip-gram models and graph neural networks, respectively.
  \item \textbf{Discrete-time network methods.} TNODE~\cite{tnode} builds a recurrent model upon node embeddings in each time step, which is a state-of-the-art method on discrete-time networks.
  \item \textbf{Continous-time network methods.} CTDNE~\cite{nguyen2018ctdne} and HTNE~\cite{zuo2018htne} are initially designed for continuous-time networks. However, they are not flexible for generating temporal node embeddings at a given timestamp. JODIE~\cite{kumar2019jodie}, TGAT~\cite{tgat_iclr20}, \zty{APAN~\cite{wang2020apan} and TGN~\cite{tgn2020rossi} are four} state-of-the-art temporal network methods.
\end{itemize}

\subsubsection{Settings of Temporal Link Prediction}
The embedding dimension for all methods is set to 128 for a fair comparison.
For each method, we report the average testing performance and standard deviation over five runs of different hyper-parameters.
For static graph methods, we use grid search of the random walk parameters $p,q$ of Node2Vec~\cite{grover2016node2vec} over $\{0.25, 0.5, 1, 2, 4\}$, and implement an inductive SAGE~\cite{hamilton2017graphsage} with uniformly sampling 100 neighbors from historical interactions.
For discrete-time graph methods, we run TNODE over three different graph snapshot divisions, namely $\{8, 32, 128\}$.
For CTDNE~\cite{nguyen2018ctdne}, the number of the walk is set at 10, the walk length is set at 80, and the context window size is set at 10.
The history length of HTNE~\cite{zuo2018htne} is searched over $\{10, 20, 30\}$.
\rtwo{The hyper-parameter settings of JODIE~\cite{kumar2019jodie}, TGAT~\cite{tgat_iclr20}, APAN~\cite{wang2020apan} and TGN~\cite{tgn2020rossi} are listed in Appendix~\ref{sec:param}.}

The proposed TIP-GNN is implemented in PyTorch~\cite{pytorch2019paszke}.
In default settings, the learning rate is set to 0.0001, and the dropout ratio is set to 0.1.
The number of \zty{TIP-GNN layers} is searched over $\{1, 2\}$, and the number of attention heads of the bilevel graph convolution is searched over $\{1, 2, 3, 4\}$.
The early-stopping strategy is adopted until the validation AUC score \zty{does not} improve for three epochs.

\subsubsection{Settings of Temporal Node Classification}
Continuous-time graph methods, including JODIE~\cite{kumar2019jodie}, APAN~\cite{wang2020apan}, TGAT~\cite{tgat_iclr20}, \zty{TGN~\cite{tgn2020rossi},} and our TIP-GNN, predict users' states using temporal node embeddings.
For other methods, we concatenate the node embeddings of the user and the corresponding Wikipedia page or subreddit as input features.
We use the three-layer MLP~\cite{tgat_iclr20} as a classifier, whose hidden dimensions are \{80,10,1\} respectively. 
In addition, the MLP classifier is trained with the Adam optimizer, the Glorot initialization, and the early-stopping strategy with ten epochs. 
\zty{Due to the data imbalance, the positive labels in each batch are oversampled for better performance.}

\begin{figure*}[t]
  \includegraphics[width=\textwidth]{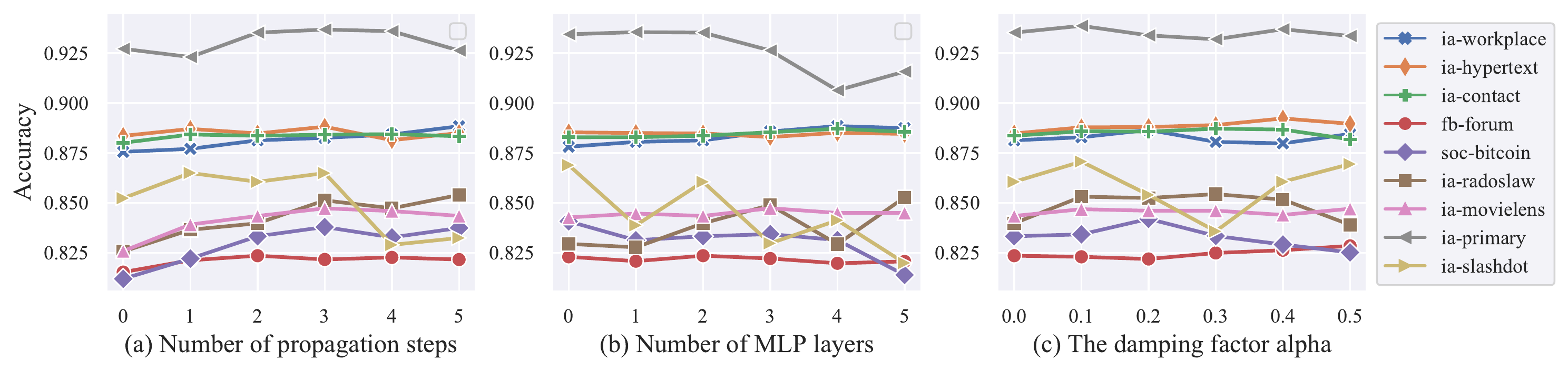}
  \caption{Ablation studies on the transition propagation module of TIP-GNN.}
  \label{fig:ablation}
\end{figure*}

\begin{figure*}[!h]
    \centering
  \includegraphics[width=.85\textwidth]{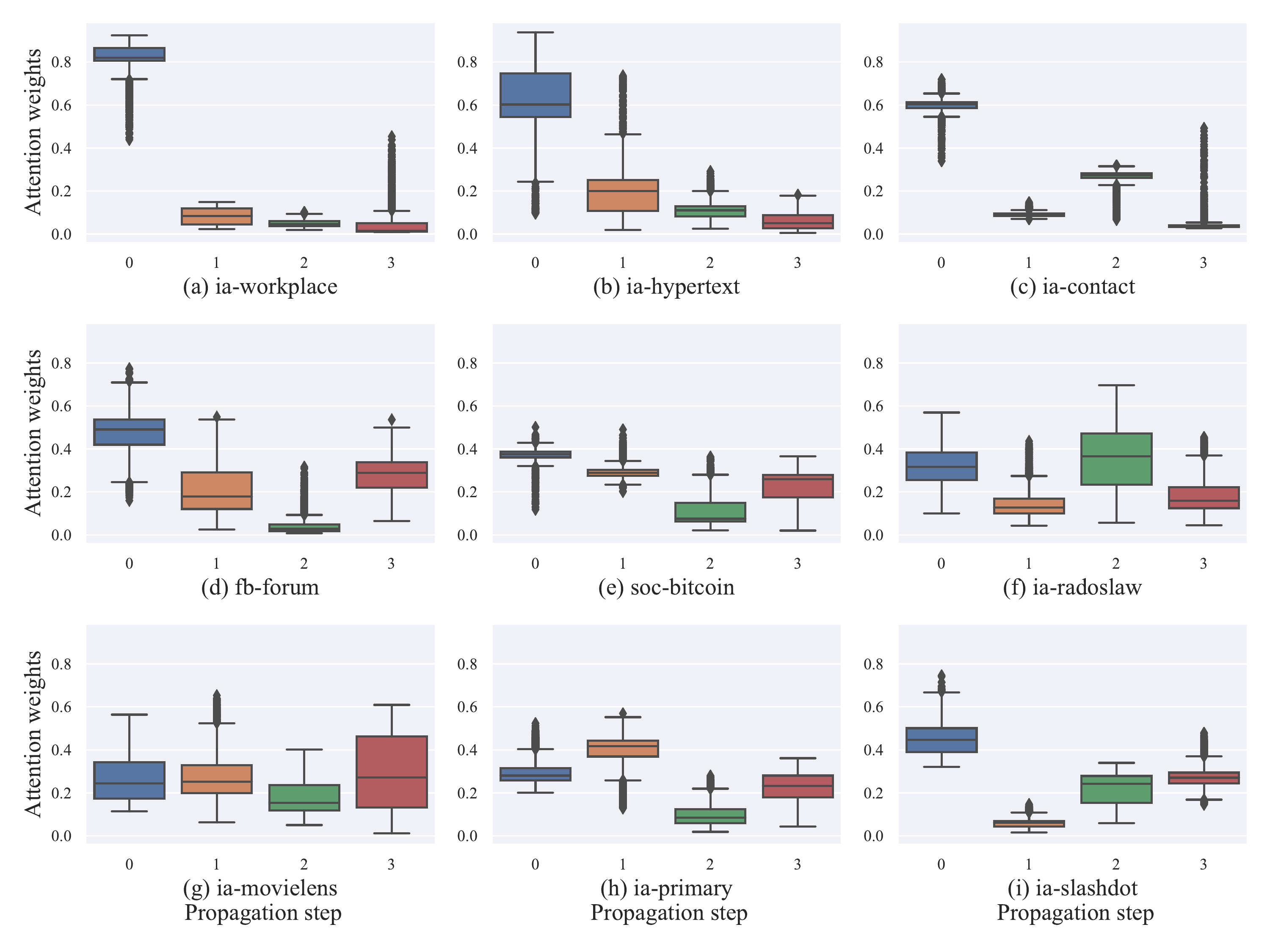}
  \caption{Attention weights at each propagation step of TIP-GNN.}
  \label{fig:visattn}
\end{figure*}

\begin{figure*}[t]
  \includegraphics[width=\textwidth]{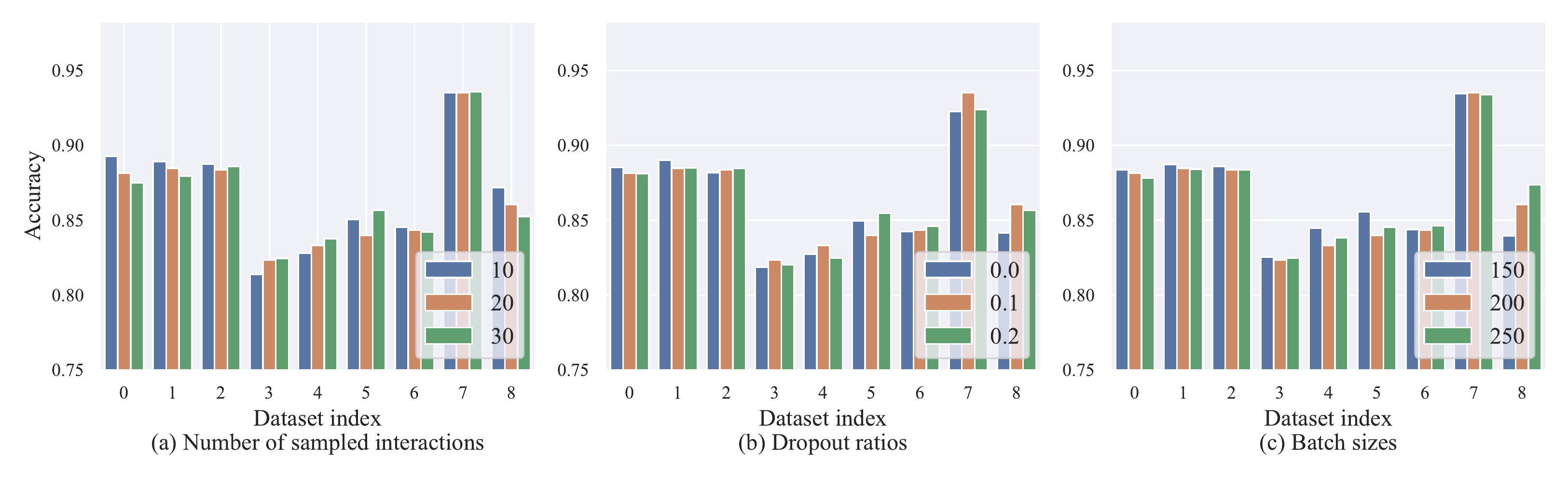}
  \caption{Parameter sensitivity of TIP-GNN.}
  \label{fig:param}
\end{figure*}

\begin{figure*}[!h]
    \centering
  \includegraphics[width=.85\textwidth]{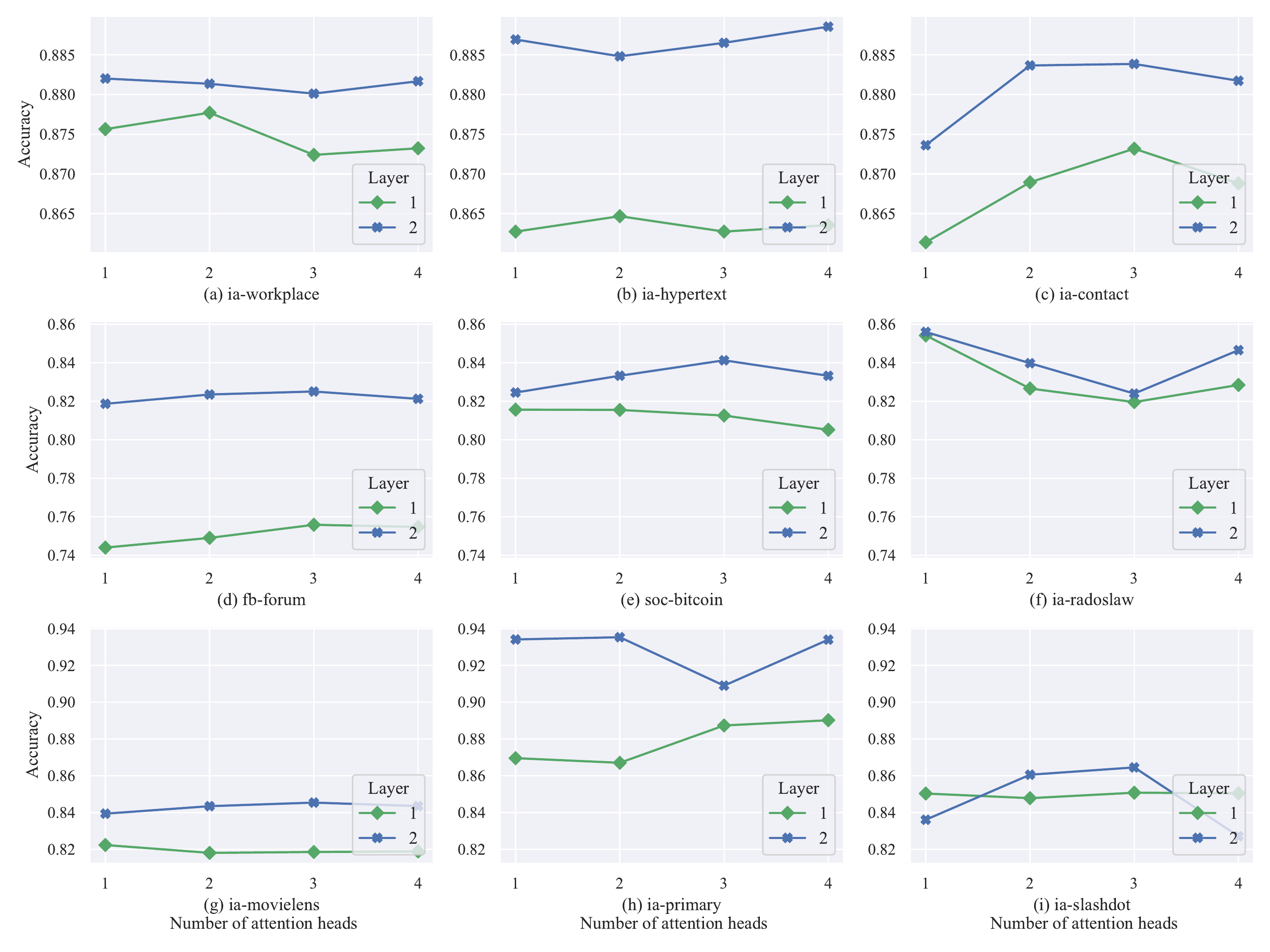}
  \caption{\zty{Effects of the number of TIP-GNN layers and attention heads.}}
  \label{fig:layer_head}
\end{figure*}
\subsection{RQ1: Overall Performance}

\subsubsection{Temporal Link Prediction}

Table~\ref{tab:auc} reports the average performance of baseline methods over five runs on temporal link prediction.
The discussions of their performance are organized as follows:
\begin{itemize}
  \item Node2Vec~\cite{grover2016node2vec} and SAGE~\cite{hamilton2017graphsage} are two robust baselines on temporal networks, although they are designed for static networks. Specifially, SAGE~\cite{hamilton2017graphsage} presents outstanding performance than CTDNE~\cite{nguyen2018ctdne}, HTNE~\cite{zuo2018htne} on the first three small and dense networks.
  \item CTDNE~\cite{nguyen2018ctdne}, HTNE~\cite{zuo2018htne}, and TNODE~\cite{tnode} are three temporal network methods. Detailedly, CTDNE performs best on large and sparse networks, while its mechanism \zty{is not} good for small and dense networks. TNODE beats HTNE on most datasets and outperforms CTDNE on small and dense networks especially.
  \item JODIE, which is initially designed for temporal item recommendation in temporal networks, demonstrates a performance decline on small and dense netowrks, namely \emph{ia-workplace}, \emph{ia-hypertext}, \emph{fb-forum}, \emph{ia-radoslaw}, and \emph{ia-primary}. Moreover, its high computational complexity causes the failed experiments of \emph{ia-slashdot} with 51,000 nodes on our P6000 GPU with a 24GB memory.
  \item APAN~\cite{wang2020apan} and TGAT~\cite{tgat_iclr20} are both based on graph neural networks. However, the dedication of APAN to inference efficiency may make it unsuitable for small networks, while TGAT achieves the second-best on most datasets. Nevertheless, TGAT shows performance degradation on the large and sparse network, namely \emph{ia-slashdot}.
  \item \zty{
    TGN~\cite{tgn2020rossi} is a lightweight and powerful temporal network method built upon the recurrent mechanism and graph neural networks.
    Table~\ref{tab:auc} shows that TGN achieves comparable AUC scores with TIP-GNN on several datasets.
    \rtwo{However, TGN underperforms TIP-GNN on \emph{fb-forum} and \emph{soc-bitcoin}, which may require the information of high-order transition structures.}
  }
  \item Our proposed TIP-GNN performs on par with or outperforms state-of-the-art methods on all datasets. Concretely, TIP-GNN and TGAT~\cite{tgat_iclr20} obtain top-2 performance on the first three networks, indicating that existing methods can well model small networks. As for large networks, TIP-GNN demonstrates consistent and significant improvements, as shown in Table~\ref{tab:auc}.
  Further, TIP-GNN shows stable performance on experimental datasets, compared with the performance degradation of JODIE~\cite{kumar2019jodie}, APAN~\cite{wang2020apan}, and TGAT~\cite{tgat_iclr20} on several networks.
  Overall, the improvements of TIP-GNN validate the importance of transition structures between neighbors.
\end{itemize}

\subsubsection{\zty{Inductive Temporal Link Prediction}}
\zty{
  Table~\ref{tab:inductive} reports the average AP scores of different methods, where most results are inherited from APAN~\cite{wang2020apan} and TGN~\cite{tgn2020rossi}.
  Overall, APAN, TGN, and TIP-GNN obtain close AP scores on \emph{Wikipedia} and \emph{Reddit}, which indicate that predicting temporal links is somewhat easy for those two datasets.
  Table~\ref{tab:data} also shows the high repetition ratios of those two datasets that our model could achieve high prediction precision by simply predicting the last neighbor.
}

\subsubsection{Temporal Node Classification} 

Table~\ref{tab:nc} reports the average AUC scores and standard deviations of methods over five runs except for TNODE~\cite{tnode} because the original node classifier of TNODE presents poor performance.
The temporal node classification task on \emph{Wikipedia} and \emph{Reddit} is challenging to all methods due to their highly imbalanced label ratios.
Specifically, static graph methods (Node2Vec~\cite{grover2016node2vec} and SAGE~\cite{hamilton2017graphsage}) outperform CTDNE and HTNE on both datasets, revealing that whether users are banned from editing Wikipedia pages or posting in the subreddits are strongly related to the target pages or subreddits.
Also, the limited memorization capacity of CTDNE and HTNE may cause poor performance on temporal node classification.
As for continuous-time network methods that can generalize to unseen nodes, JODIE~\cite{kumar2019jodie}, APAN~\cite{wang2020apan}, TGAT~\cite{tgat_iclr20}, \zty{TGN~\cite{tgn2020rossi}}, and our TIP-GNN all demonstrate much better performance than other methods.
\zty{
  Specifically, APAN~\cite{wang2020apan} performs best on \emph{Wikipedia} and TGN~\cite{tgn2020rossi} achieves the second-best on \emph{Reddit}, indicating the dataset preferences of models.
}
On \emph{Reddit}, TIP-GNN performs significantly better than other methods, which may be attributed to the increased density of \emph{Reddit} than \emph{Wikipedia}.

\subsection{RQ2: Ablation Study}
\label{sec:ablation}

\subsubsection{Designs of the transition propagation module}

The transition propagation module, the core of TIP-GNN, contains variable propagation steps, the nonlinear transformation MLP, and the damping factor $\alpha$ for personalized embedding.
In default settings, the number of propagation steps is 2, MLP layers are 2, and the damping factor $\alpha$ is 0.
As shown in Fig.~\ref{fig:ablation}, the effects of these designs are summarized as follows:
\begin{itemize}
  \item The propagation at different steps aims at representing dynamic patterns at different granularities. Fig.~\ref{fig:ablation}(a) shows the obvious improvements of accuracy with the increase of propagation steps. However, the performance on most datasets achieves its peak at three propagation steps. It may be caused by aggregated noise from distant neighborhoods and the well-known over-smoothing problem of GNNs~\cite{kipf2016semi}.
  \item The designed MLP aims at enhancing the nonlinearity of embeddings during transition propagation. On the contrary, Fig.~\ref{fig:ablation}(b) shows that the performance is almost irrelevant to the MLP layers and even suffers from a deep MLP. Significantly, the 0-layer MLP, which propagates embeddings without any parametrized projection, achieves comparable performance to the best performance on most datasets. This finding could be attributed to (1) the additional nonlinearity of the bilevel graph convolution module and (2) the uselessness of nonlinear transformation during graph convolution~\cite{sgcn2019wu,lightgcn2020he}.
  \item The damping factor $\alpha$~\cite{ppnp2019klicpera} works as a residual connection to overcome the over-smoothing problem and the training difficulty of GNNs. Fig.~\ref{fig:ablation}(c) shows complex relations between $\alpha$ and the performance: social networks with comments and replies (i.e., \emph{ia-radoslaw} and \emph{ia-slashdot}) benefit significantly from the damping factor, while other networks perform stably regardless of their sparsity and repetition ratios. In general, a small damping factor could boost the performance, but an optimal damping factor depends on the network properties.
\end{itemize}

\subsubsection{Discussions on the propagation step}

The propagation step defined by the Eq.~\ref{eq:prop-embed} plays an essential role in the transition propagation module, where the number of MLP layers is 2 and the damping factor $\alpha$ is 0.
Since most datasets achieve the peak performance when the propagation step is 3 in Fig.~\ref{fig:ablation}(a), we visualize the corresponding attention weights of each propagation step defined by Eq.~\ref{eq:atte-fusion} in the last \zty{TIP-GNN} layer.
Findings from Fig.~\ref{fig:visattn} can be concluded as follows:
\begin{itemize}
  \item The 0-th step, which has not yet involved any transition propagation, usually obtains the most significant weights on most datasets compared with other propagation steps. Specifically, attention weights of the 0-th step are larger than 0.6 average on the first three networks, indicating that transition propagation may only boost the performance marginally for simple networks.
  \item With the increase of interactions, attention weights of the 0-th step decrease accordingly, and attention weights of other steps even present on par with the 0-th step on several networks, namely \emph{ia-radoslaw}, \emph{ia-movielens}, and \emph{ia-primary}. This observation validates the existence and benefits of transition structures in temporal networks. 
  \item Lastly, attention weights of different temporal networks present various distributions of propagation steps. For example, \emph{soc-bitcoin} and \emph{ia-slashdot} are both large and sparse networks. However, \emph{soc-bitcoin} prefers short-range transitions, while \emph{ia-slashdot} prefers long-range transitions. Nevertheless, our proposed attention fusion module could learn preferences adaptively across different propagation steps.
\end{itemize}

\subsection{RQ3: Parameter Sensitivity}

For quantitative comparison, the following experiments are conducted with fixed parameters of the transition propagation module that the number of propagation steps is 2, the number of MLP layers is 2, and the damping factor $\alpha$ is 0.

\subsubsection{Model Architecture}

The parameters of model architecture refer to the \zty{TIP-GNN} layers and the attention heads of the transition pooling module.
The training parameters are also fixed for simplicity that the number of \zty{sampled interactions} is 20, the dropout ratio is 0.1, and the batch size is 200.
We plot the grid search results of the number of \zty{TIP-GNN} layers over $\{1, 2\}$ and the number of attention heads over $\{1, 2, 3, 4\}$ in Fig.~\ref{fig:layer_head}, which can be summarized as follows:
\begin{itemize}
  \item 
  A deeper \zty{TIP-GNN} layer perceives a wider neighborhood in networks. 
  Fig.~\ref{fig:layer_head} demonstrates the superiority of a deeper layer on all networks, owing to the benefits of collaborative filtering~\cite{wang2019ngcf,lightgcn2020he} and neighborhood smoothing~\cite{kipf2016semi,sgcn2019wu}.
  \item 
  A larger number of attention heads enlarges the model capacity~\cite{velivckovic2017gat,attention2017vaswani}. 
  The results in Fig.~\ref{fig:layer_head} can be divided into two kinds: networks with low-repetition ratios in Table~\ref{tab:data} benefit from more attention heads, while networks with high-repetition ratios (such as \emph{ia-radoslaw} and \emph{ia-primary}) suffer from it on the contrary. 
  It indicates that a complex model may degrade the performance of simple networks.
\end{itemize}

\subsubsection{Hyper-parameters}

The compared hyper-parameters presented here include the number of neighbors, the dropout ratios, and the batch sizes.
The parameters of model architectures are fixed that the number of \zty{TIP-GNN} layers is 2, and the number of attention heads is 2.
The first three networks present stable performance with different hyper-parameters.
Thus, the following discussions only care about the last six networks:
\begin{itemize}  
  \item 
  Interactions with recent neighbors usually reveal nodes' dynamic patterns in temporal networks. 
  Fig~\ref{fig:param}(a) shows that most networks benefit from a larger window of temporal neighbors, while \emph{ia-slashdot} presents the contrary results.
  \zty{
  Fig~\ref{fig:param}(a) shows that most networks benefit from sampling more latest interactions, while \emph{ia-slashdot} presents the contrary results.}
  The reason may be that nodes in most networks demonstrate long-range interests, while \emph{ia-slashdot} demonstrate fast shifts of node interests because it is a post-then-reply network.
  \item 
  Dropout is effective for training a robust neural network model by setting some neurons of the model as zeros and adjusting the outputs accordingly~\cite{dropout2012hinton}
  Fig.~\ref{fig:param}(b) shows that our original TIP-GNN is robust compared with training under the dropout technique.
  Nevertheless, TIP-GNN achieves slight performance improvements when using dropout.
  \item
  It is a \emph{de facto} standard to train neural network models with mini-batches.
  Similar to the effects of the dropout technique, Fig.~\ref{fig:param}(c) shows the robustness of our TIP-GNN on most networks under different batch sizes.
  Specifically, the network \emph{ia-slashdot} benefits from a larger batch size, which may be attributed to its large-scale nodes and sparsity of interactions.
\end{itemize}

\section{Conclusion}

In this paper, we propose a novel TIP-GNN based on graph neural networks to model the bilevel graph structure in temporal networks, which highlights nodes' personalized transition patterns among neighbors.
The transition propagation module encodes nodes' dynamic patterns with personalized propagation rules; then, the \zty{bilevel} graph convolution module extracts useful neighborhood embeddings with a bilevel attention mechanism.
Experimental results demonstrate the superiority of encoding dynamic patterns in temporal networks that our TIP-GNN outperforms existing approaches consistently and significantly on experimental datasets.
Extensive experiments further reveal the adaptivity of the multistep propagation of TIP-GNN for various temporal networks, as well as the robustness of TIP-GNN.

As for future works, on the one hand, we will explore the representation limitation of GNNs to overcome the performance decline in Fig~\ref{fig:ablation}(a).
Our ablation studies on propagation steps shown in Fig.~\ref{fig:ablation}(a) demonstrate that TIP-GNN will achieve peak performance when enlarging the number of propagation steps properly. 
Following works towards mitigating the over-smoothing problem of GNNs~\cite{jknet2019xu,ppnp2019klicpera}, we additionally design controllable modules with the zero-layer MLP and the attention mechanism for different propagation steps.
However, the bilevel attention mechanism \zty{cannot} suppress additional noise induced by a larger propagation step.
In the future, we will investigate the over-smoothing problem from the perspective of temporal networks and propose an enhanced GNN.

On the other hand, we will explore the interpretability of our proposed transition graphs following previous works~\cite{igmc2020zhang,intents2021wang}.
IGMC~\cite{igmc2020zhang} reveals the common patterns of enclosing subgraphs between users and items, while KGIN~\cite{intents2021wang} captures user intents with an attentive mechanism over Knowledge graph relations.
In the future, we will investigate how to annotate user interactions with auxiliary information like public Knowledge Graphs.

\section{Acknowledgment}

This work is supported by National Natural Science Foundation of China (U20B2066, 61976186), the Starry Night Science Fund of Zhejiang University Shanghai Institute for Advanced Study (Grant No. SN-ZJU-SIAS-001)
, and the Fundamental Research Funds for the Central Universities (226-2022-00064).

\ifCLASSOPTIONcaptionsoff
  \newpage
\fi

\bibliographystyle{IEEEtran}
\bibliography{ref}

\newpage
\begin{IEEEbiography}
	[{\includegraphics[width=1in,height=1.25in,clip,keepaspectratio]{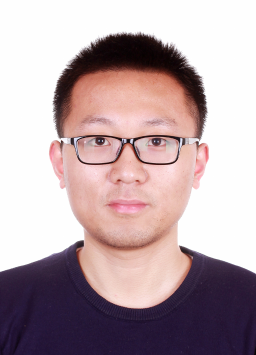}}]
	{Tongya Zheng}
	is currently pursuing the Ph.D degree with the College of Computer Science, Zhejiang University. His research interests include stream data computing and graph mining.
\end{IEEEbiography}
\vskip -2\baselineskip plus -1fil

\begin{IEEEbiography}
	[{\includegraphics[width=1in,height=1.25in,clip,keepaspectratio]{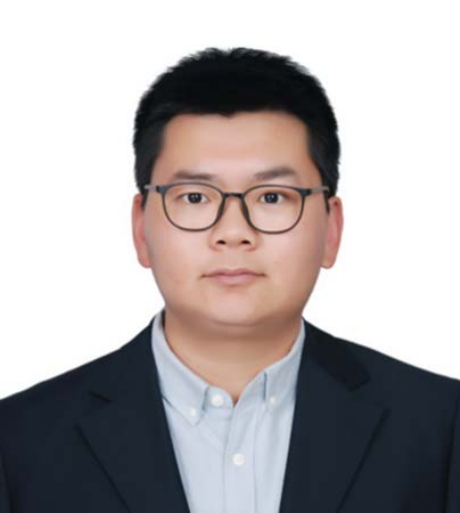}}]
	{Zunlei Feng}
	is an assistant research fellow in College of Software Technology, Zhejiang University. He received his Ph.D degree in Computer Science and Technology from College of Computer Science, Zhejiang University, and B. Eng. Degree from Soochow University. His research interests mainly include computer vision, image information processing, representation learning, medical image analysis. He has authored and co-authored many scientific articles at top venues including IJCV, NeurIPS, AAAI, TVCG, ACM TOMM, and ECCV. He has served with international conferences including AAAI and PKDD, and international journals including IEEE Transactions on Circuits and Systems for Video Technology, Information Sciences, Neurocomputing, Journal of Visual Communication and Image Representation and Neural Processing Letters.
\end{IEEEbiography}
\vskip -2\baselineskip plus -1fil

\vskip -2\baselineskip plus -1fil
\begin{IEEEbiography}
	[{\includegraphics[width=1in,height=1.25in,clip,keepaspectratio]{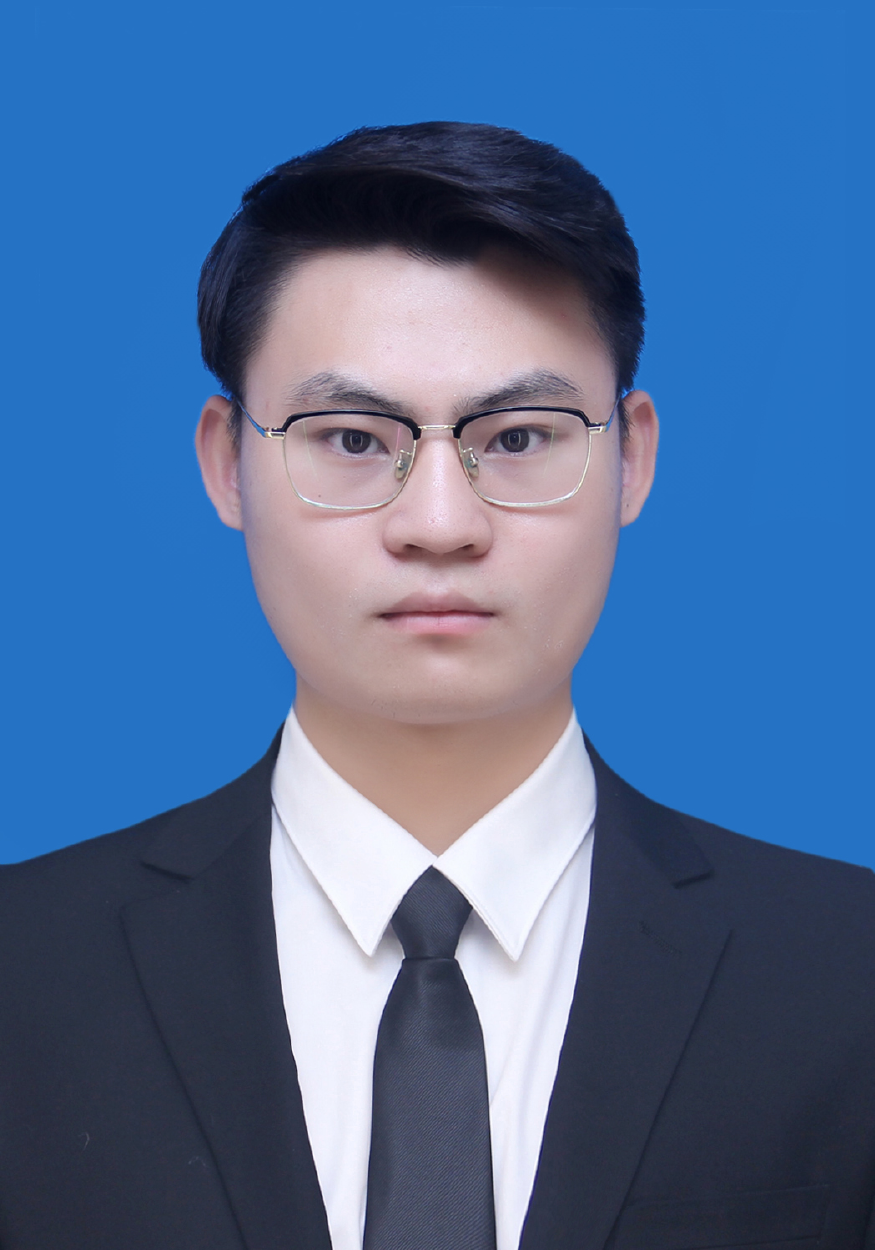}}]
	{Tianli Zhang}
	is currently pursuing the M.D degree with the College of Software Technology, Zhejiang University. His research interests include graph mining.
\end{IEEEbiography}
\vskip -2\baselineskip plus -1fil

\vskip -2\baselineskip plus -1fil
\begin{IEEEbiography}
	[{\includegraphics[width=1in,height=1.25in,clip,keepaspectratio]{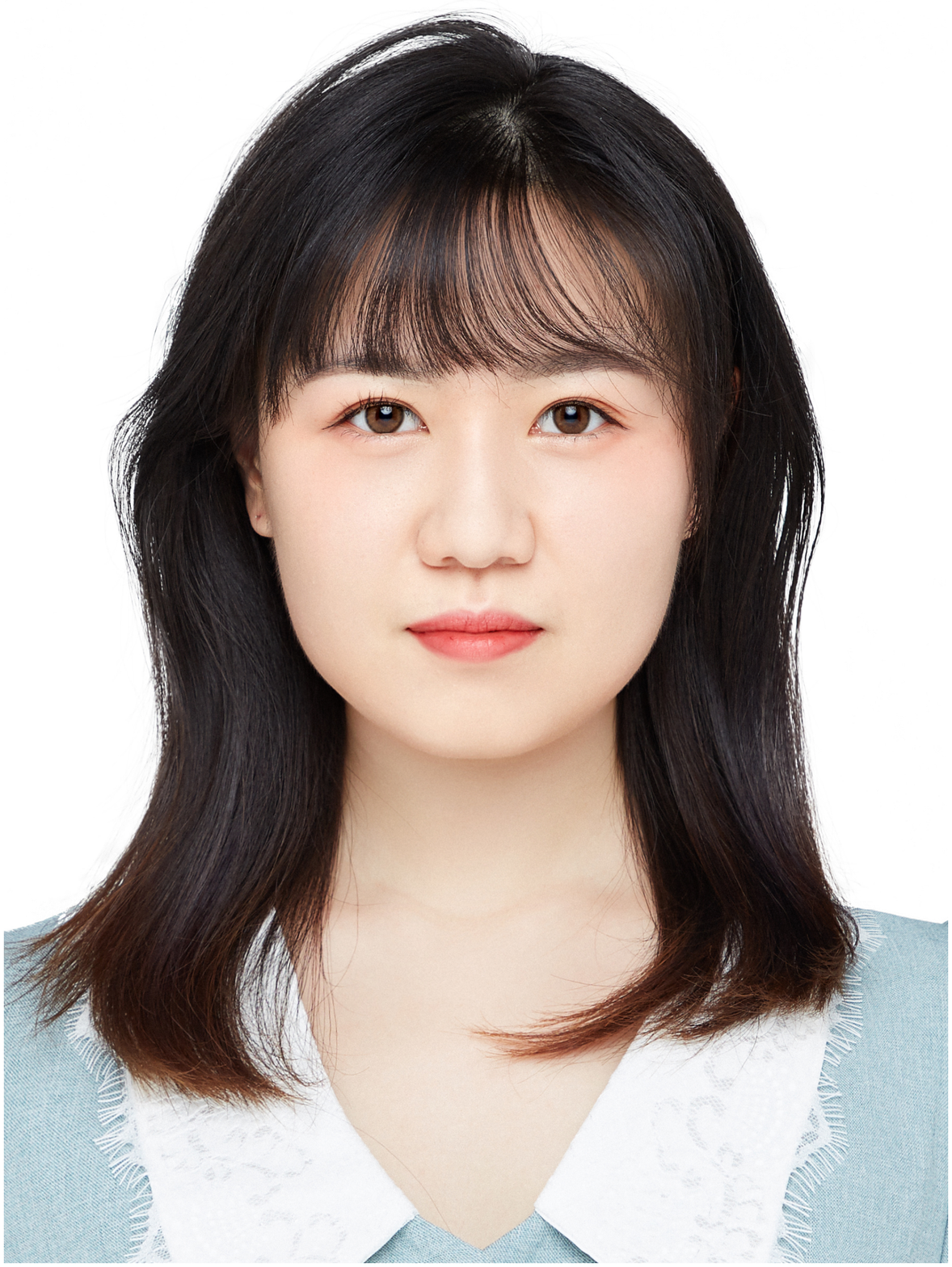}}]
	{Yunzhi Hao}
	is currently pursuing the Ph.D degree with the College of Computer Science, Zhejiang University. Her research interests include graph inference and network embedding.
\end{IEEEbiography}
\vskip -2\baselineskip plus -1fil

\begin{IEEEbiography}
	[{\includegraphics[width=1in,height=1.25in,clip,keepaspectratio]{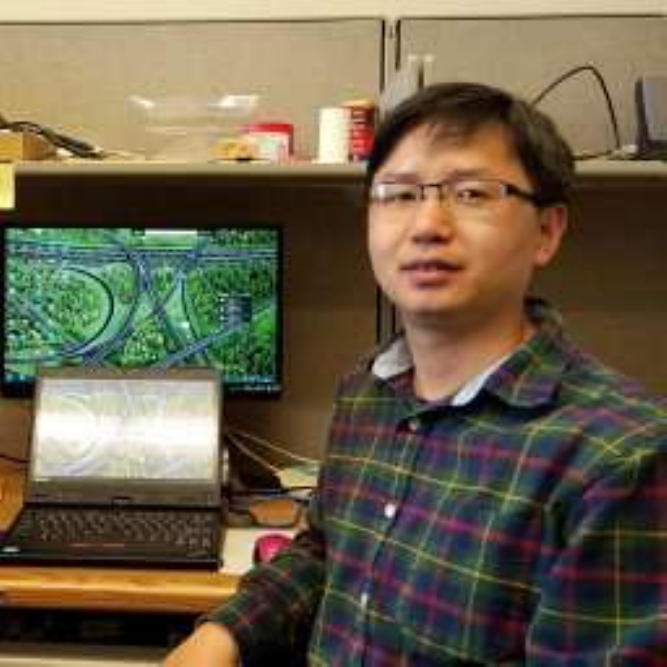}}]
	{Mingli Song}
	received the Ph.D degree in computer science from Zhejiang University, China, in 2006. He is currently a Professor with the Microsoft Visual Perception Lab- oratory, Zhejiang University. His research interests include face modeling and facial expression analysis. He received the Microsoft Research Fellowship in 2004.
\end{IEEEbiography}
\vskip -2\baselineskip plus -1fil

\begin{IEEEbiography}
	[{\includegraphics[width=1in,height=1.25in,clip,keepaspectratio]{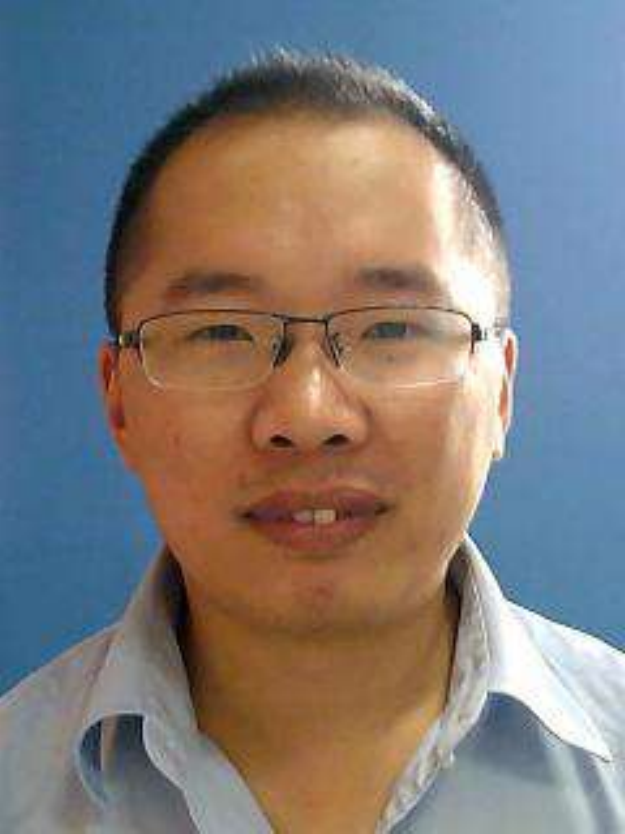}}]
	{Xingen Wang}
	received the graduate and Ph.D degrees in computer science from Zhejiang University of China, in 2005 and 2013, respectively. He is currently a research assistant in the College of Computer Science, Zhejiang University. His research interests include distributed computing and software performance.
\end{IEEEbiography}
\vskip -2\baselineskip plus -1fil

\begin{IEEEbiography}
	[{\includegraphics[width=1in,height=1.25in,clip,keepaspectratio]{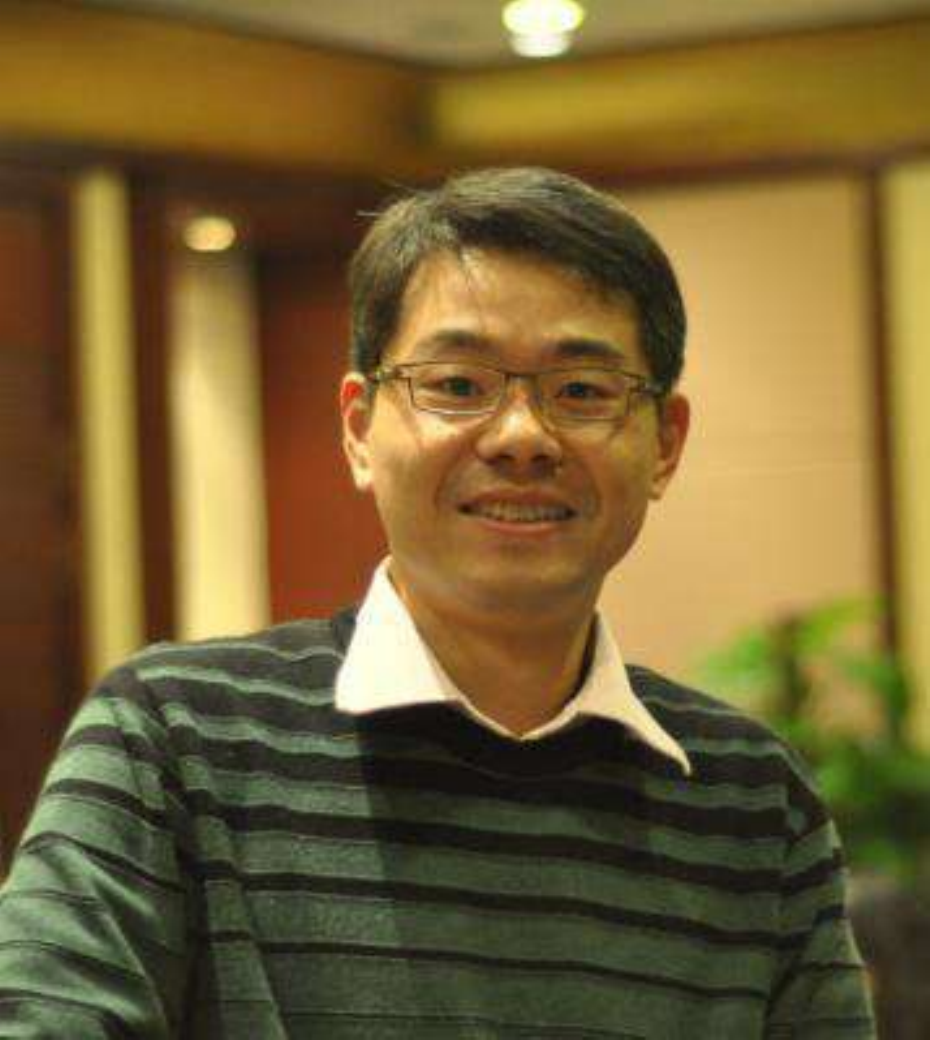}}]
	{Xinyu Wang}
	received the graduate and Ph.D degrees in computer science from Zhejiang University of China, in 2002 and 2007, respectively. He was a research assistant at the Zhejiang University, from 2002 to 2007. He is currently a professor in the College of Computer Science, Zhejiang University. His research interests include streaming data analysis, formal methods, very large information systems, and software engineering.
\end{IEEEbiography}
\vskip -2\baselineskip plus -1fil

\begin{IEEEbiography}
	[{\includegraphics[width=1in,height=1.25in,clip,keepaspectratio]{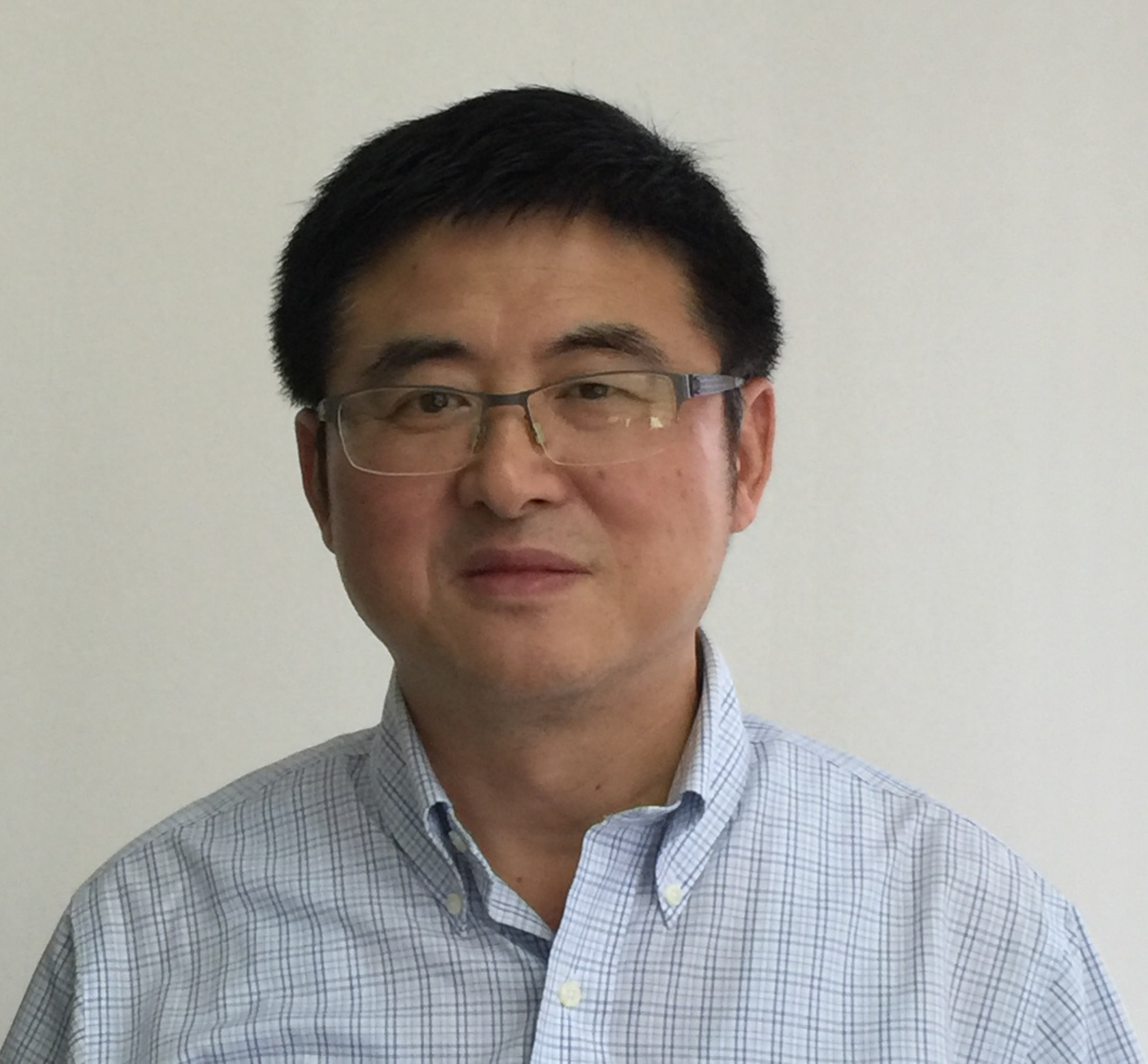}}]
	{Ji Zhao}
	received the Ph.D. degree in computer science from Zhejiang University, China. His research interests include data mining, financial informatization and fintech. He is a review expert of Shanghai government procurement, the head of information Technology risk Team of Shanghai Banking Network Industry Association, and has successively served for Shanghai Pudong Development Bank Shanghai Branch.
\end{IEEEbiography}
\vskip -2\baselineskip plus -1fil

\begin{IEEEbiography}
	[{\includegraphics[width=1in,height=1.25in,clip,keepaspectratio]{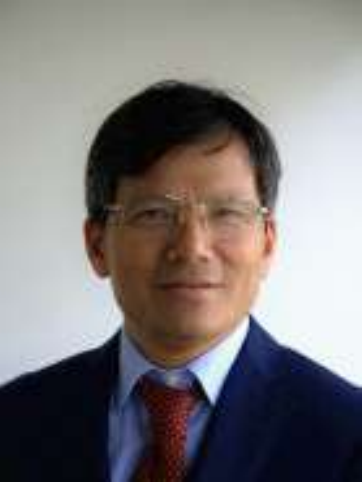}}]
	{Chun Chen}
	is currently a Professor with the College of Computer Science, Zhejiang University. His research interests include computer vision, computer graphics, and embedded technology.
\end{IEEEbiography}

\clearpage

\appendix

\section{Appendices}
\subsection{Inductive experiments for datasets without features}
\label{sec:inductive}

\rtwo{For temporal networks without node or edge features, we firstly obtain classical node-level features based on the graph topology using NetworkX~\footnote{https://networkx.org/documentation/stable/reference/algorithms/centrality.html}, which are sixty dimensions in total.
The temporal network with multiple edges between the same node pair is transformed into a weighted simple graph before computing node features.
The node features include ten dimensions of node centrality metrics and the first fifty dimensions of eigenvectors of the laplacian matrix, listed as follows:
\begin{itemize}
  \item degree centrality, closeness centrality, betweenness centrality, load centrality, PageRank centrality, clustering coefficient, number of triangles, hub centrality, authority centrality, and laplacian spectrum;
  \item and the first fifty dimensions of eigenvectors of the laplacian matrix.
\end{itemize}
New nodes in the validation and testing set also obtain their node features in this way.}

\rtwo{
For a fair comparison, we have tested the hyper-parameters of comparison methods according to their ablation studies and searched the same training hyper-parameters as the proposed TIP-GNN, as shown in Table~\ref{tab:params}.
The experimental results are summarized in Table~\ref{tab:inductive-fb}.
Firstly, the performance change of TIP-GNN (-3.8\%$\sim$6.3\%) is much more stable than existing methods based on different node features, namely SAGE (-13.5\%$\sim$26.0\%), JODIE (-8.2\%$\sim$35.4\%), TGAT (-19.9\%$\sim$10.6\%), and TGN (-10.1\%$\sim$9.8\%).
Secondly, TIP-GNN could achieve better performance than existing methods consistently on seven datasets both in the transductive setting and the inductive setting.
Although TIP-GNN underperforms TGN~\cite{tgn2020rossi} on \emph{ia-contact}, TIP-GNN also achieves the second-best.
Thirdly, it is noteworthy that sparse networks like \emph{soc-bitcoin, ia-movielens, and ia-slashdot} mostly benefit from new features, and other dense networks mostly suffer from new features.
This may be caused by that nodes in sparse networks are not trained sufficiently and are improved by new features, while nodes in dense networks usually perform repeated interactions with the same neighbor, which are ignored by new features.}

\subsection{Hyper-parameters of baseline methods}
\label{sec:param}

\rtwo{The hyper-parameter settings of JODIE~\cite{kumar2019jodie}, TGAT~\cite{tgat_iclr20}, APAN~\cite{wang2020apan} and TGN~\cite{tgn2020rossi} are listed in Table~\ref{tab:params}.}

\begin{table}[t]
  \caption{Hyper-parameters of baseline methods.}
  \centering 
  \resizebox{\linewidth}{!}{
  \begin{tabular}{l|l}
    \toprule
    Method & Parameter \\
    \midrule
    JODIE~\cite{kumar2019jodie} & Epoch $\in \{1, \cdots, 50\}$ \\
    \midrule
    \multirow{4}{*}{TGAT~\cite{tgat_iclr20}} & Layer $\in \{1, 2\}$ \\
    & Number of attention heads $\in \{1, 2, 3, 4\}$ \\ 
    & Sampler $\in$ \{uniform, temporal\}  \\
    & Time encoding $\in$ \{harmonic, position, empty\}  \\
    & Number of sampled neighbors $\in \{10, 20, 30\}$  \\
    & Dropout ratio $\in \{0.0, 0.1, 0.2\}$  \\
    & Batch size $\in \{150, 200, 250\}$  \\
    \midrule
    \multirow{3}{*}{APAN~\cite{wang2020apan}} & Layer $\in \{1, 2\}$  \\
    & Number of attention heads $\in \{1, 2, 3, 4\}$  \\
    & Sampler $\in$ \{uniform, temporal\}  \\
    & Number of sampled neighbors $\in \{10, 20, 30\}$  \\
    & Dropout ratio $\in \{0.0, 0.1, 0.2\}$  \\
    & Batch size $\in \{150, 200, 250\}$  \\
    \midrule
    \multirow{5}{*}{TGN~\cite{tgn2020rossi}} & Layer $\in \{1, 2\}$  \\
    & Aggregator $\in$ \{last, mean\}  \\
    & Number of attention heads $\in \{1, 2, 3, 4\}$  \\
    & Embedding $\in$ \{attention, sum, identity, time\} \\ 
    & Sampler $\in$ \{uniform, temporal\}  \\
    & Number of sampled neighbors $\in \{10, 20, 30\}$  \\
    & Dropout ratio $\in \{0.0, 0.1, 0.2\}$  \\
    & Batch size $\in \{150, 200, 250\}$  \\
    \bottomrule
  \end{tabular}
  }
  \label{tab:params}
\end{table}

\begin{table*}[!h]
  \caption{\rtwo{AUC scores of transductive and inductive temporal link prediction. 
  Bold font indicates the best performance.
  The inductive learning methods (denoted by *) based on new features are compared against their performance based on one-hot features to verify their robustness with different node features.
  $\redarrow$ refers to the performance improvement, and $\greenarrow$ refers to the performance decline.
  The symbol $-$ is used for failed experiments.
  }
  }
  \centering
  \resizebox{1.0\textwidth}{!}{
  \begin{tabular}{lcccccc}
    \toprule
    & \multicolumn{2}{c}{ \textbf{ ia-workplace }} & \multicolumn{2}{c}{ \textbf{ ia-hypertext }} & \multicolumn{2}{c}{ \textbf{ ia-contact }} \\
    \cmidrule(){2-7}
		& Transductive & Inductive  & Transductive & Inductive  & Transductive & Inductive \\
		\midrule
    SAGE & $ 0.857 \pm 0.008 $ & $ - $  & $ 0.830 \pm 0.006 $ & $ - $  & $ 0.856 \pm 0.003 $ & $ - $ \\
    SAGE* & $ 0.810 \pm 0.011 \greenarrow $ & $ 0.816 \pm 0.058 $ & $ 0.745 \pm 0.033 \greenarrow$ & $ 0.685 \pm 0.032 $ & $ 0.902 \pm 0.006 \redarrow$ & $ 0.496 \pm 0.035 $\\
    Change & -5.4\% & & -10.2\% & & 5.4\% & \\
		\midrule
    JODIE & $ 0.600 \pm 0.054 $ & $ - $  & $ 0.667 \pm 0.067 $ & $ - $  & $ 0.850 \pm 0.005 $ & $ - $ \\
    JODIE* & $ 0.618 \pm 0.012\redarrow $ & $ 0.812 \pm 0.082 $ & $ 0.778 \pm 0.012 \redarrow$ & $ 0.664 \pm 0.026 $ & $ 0.920 \pm 0.004 \redarrow$ & $ 0.612 \pm 0.117 $\\
    Change & 3.0\% & & 16.7\% & & 8.2\% & \\
		\midrule
    TGAT & $ 0.959 \pm 0.001 $ & $ - $  & $ \bm{0.959 \pm 0.001} $ & $ - $  & $ 0.921 \pm 0.001 $ & $ - $ \\
    TGAT* & $ 0.852 \pm 0.105 \greenarrow$ & $ 0.653 \pm 0.000 $ & $ 0.806 \pm 0.132 \greenarrow$ & $ 0.701 \pm 0.114 $ & $ 0.907 \pm 0.009 \greenarrow$ & $ 0.487 \pm 0.004 $\\
    Change & -0.8\% & & -16.0\% & & -1.5\% & \\
		\midrule
    TGN & $ 0.960 \pm 0.001 $ & $ - $  & $ 0.952 \pm 0.001 $ & $ - $  & $ 0.927 \pm 0.001 $ & $ - $ \\
    TGN* & $ 0.950 \pm 0.010 \greenarrow$ & $ 0.852 \pm 0.031 $ & $ 0.908 \pm 0.015 \greenarrow$ & $ 0.829 \pm 0.030 $ & $ 0.928 \pm 0.003 \redarrow$ & $ \bm{0.724 \pm 0.034} $\\
    Change & -1.0\% & & -4.6\% & & 0.1\% & \\
		\midrule
    TIP-GNN & $ 0.961 \pm 0.007 $ & $ - $  & $ 0.958 \pm 0.005 $ & $ - $  & $ 0.922 \pm 0.003 $ & $ - $ \\
    TIP-GNN* & $ \bm{ 0.963 \pm 0.002 \redarrow} $ & $ \bm{ 0.857 \pm 0.020 } $ & $ { 0.956 \pm 0.002\greenarrow } $ & $ \bm{ 0.867 \pm 0.013 } $ & $ \bm{ 0.947 \pm 0.001 \redarrow} $ & $ 0.645 \pm 0.027 $\\
    Change & 0.2\% & & -0.2\% & & 2.7\% & \\
    \toprule
    & \multicolumn{2}{c}{\textbf{fb-forum}} & \multicolumn{2}{c}{\textbf{soc-bitcoin}} & \multicolumn{2}{c}{\textbf{ia-radoslaw}} \\  
    \cmidrule(){2-7}
		& Transductive & Inductive  & Transductive & Inductive  & Transductive & Inductive \\
		\midrule
    SAGE & $ 0.724 \pm 0.010 $ & $ - $  & $ 0.734 \pm 0.009 $ & $ - $  & $ 0.894 \pm 0.005 $ & $ - $ \\
    SAGE* & $ 0.791 \pm 0.006 \redarrow$ & $ 0.800 \pm 0.007 $ & $ 0.925 \pm 0.010 \redarrow$ & $ 0.732 \pm 0.009 $ & $ 0.789 \pm 0.016 \greenarrow$ & $ 0.500 \pm 0.500 $\\
    Change & 9.3\% & & 26.0\% & & -11.7\% & \\
		\midrule
    JODIE & $ 0.751 \pm 0.045 $ & $ - $  & $ 0.880 \pm 0.019 $ & $ - $  & $ 0.784 \pm 0.022 $ & $ - $ \\
    JODIE* & $ 0.685 \pm 0.050 \greenarrow$ & $ 0.665 \pm 0.073 $ & $ 0.859 \pm 0.042 \greenarrow$ & $ 0.767 \pm 0.156 $ & $ 0.770 \pm 0.038 \greenarrow$ & $ 0.800 \pm 0.447 $\\
    Change & -8.7\% & & -2.4\% & & -1.8\% & \\
		\midrule
    TGAT & $ 0.877 \pm 0.001 $ & $ - $  & $ 0.896 \pm 0.001 $ & $ - $  & $ 0.919 \pm 0.001 $ & $ - $ \\
    TGAT* & $ 0.796 \pm 0.021 \greenarrow$ & $ 0.782 \pm 0.001 $ & $ 0.930 \pm 0.000 \redarrow$ & $ 0.801 \pm 0.018 $ & $ 0.779 \pm 0.005 \greenarrow$ & $ 1.000 \pm 0.000 $\\
    Change & -9.2\% & & 3.8\% & & -15.2\% & \\
		\midrule
    TGN & $ 0.889 \pm 0.016 $ & $ - $  & $ 0.838 \pm 0.004 $ & $ - $  & $ 0.885 \pm 0.008 $ & $ - $ \\
    TGN* & $ 0.833 \pm 0.022 \greenarrow$ & $ 0.814 \pm 0.025 $ & $ 0.920 \pm 0.009 \redarrow$ & $ 0.791 \pm 0.024 $ & $ 0.816 \pm 0.002 \greenarrow$ & $ 0.400 \pm 0.516 $\\
    Change & -6.3\% & & 9.8\% & & -7.8\% & \\
		\midrule
    TIP-GNN & $ \bm{0.911 \pm 0.006} $ & $ - $  & $ 0.912 \pm 0.006 $ & $ - $  & $ \bm{0.929 \pm 0.005} $ & $ - $ \\
    TIP-GNN* & $ { 0.876 \pm 0.002 \greenarrow} $ & $ \bm{ 0.858 \pm 0.003 } $ & $ \bm{ 0.969 \pm 0.001 \redarrow} $ & $ \bm{ 0.921 \pm 0.003 } $ & $ { 0.927 \pm 0.002 \greenarrow} $ & $ \bm{ 1.000 \pm 0.000 } $\\
    Change & -3.8\% & & 6.3\% & & -0.2\% & \\
    \toprule
    & \multicolumn{2}{c}{\textbf{ia-movielens}} & \multicolumn{2}{c}{\textbf{ia-primary}} & \multicolumn{2}{c}{\textbf{ia-slashdot}}\\  
		\cmidrule(){2-7}
		& Transductive & Inductive  & Transductive & Inductive  & Transductive & Inductive \\
		\midrule
    SAGE & $ 0.800 \pm 0.004 $ & $ - $  & $ 0.929 \pm 0.001 $ & $ - $  & $ 0.782 \pm 0.005 $ & $ - $ \\
    SAGE* & $ 0.905 \pm 0.008 \redarrow$ & $ 0.809 \pm 0.038 $ & $ 0.803 \pm 0.014 \greenarrow$ & $ - $ & $ 0.832 \pm 0.009 \redarrow$ & $ 0.649 \pm 0.012 $\\
    Change & 13.1\% & & -13.5\% & & 6.4\% & \\
		\midrule
    JODIE & $ 0.862 \pm 0.006 $ & $ - $  & $ 0.588 \pm 0.012 $ & $ - $  & $ - $ & $ - $ \\
    JODIE* & $ 0.830 \pm 0.016 \greenarrow$ & $ 0.787 \pm 0.063 $ & $ 0.796 \pm 0.040 \redarrow$ & $ - $ & $ - $ & $ - $\\
    Change & -3.7\% & & 35.4\% & & & \\
		\midrule
    TGAT & $ 0.884 \pm 0.001 $ & $ - $  & $ 0.951 \pm 0.001 $ & $ - $  & $ 0.805 \pm 0.001 $ & $ - $ \\
    TGAT* & $ 0.913 \pm 0.002 \redarrow$ & $ 0.855 \pm 0.002 $ & $ 0.762 \pm 0.001 \greenarrow $ & $ - $ & $ 0.890 \pm 0.001 \redarrow$ & $ 0.752 \pm 0.001 $\\
    Change & 3.2\% & & -19.9\% & & 10.6\% & \\
		\midrule
    TGN & $ 0.921 \pm 0.001 $ & $ - $  & $ 0.961 \pm 0.011 $ & $ - $  & $ 0.896 \pm 0.005 $ & $ - $ \\
    TGN* & $ 0.927 \pm 0.005 \redarrow$ & $ 0.895 \pm 0.009 $ & $ 0.864 \pm 0.013 \greenarrow$ & $ - $ & $ 0.919 \pm 0.006 \redarrow$ & $ 0.860 \pm 0.068 $\\
    Change & 6.5\% & & -10.1\% & & 2.6\% & \\
		\midrule
    TIP-GNN & $ 0.922 \pm 0.002 $ & $ - $  & $ \bm{0.978 \pm 0.001} $ & $ - $  & $ 0.938 \pm 0.004 $ & $ - $ \\
    TIP-GNN* & $ \bm{ 0.967 \pm 0.001 \redarrow} $ & $ \bm{ 0.937 \pm 0.014 } $ & $ { 0.977 \pm 0.001 \greenarrow} $ & $ - $ & $ \bm{ 0.939 \pm 0.001 \redarrow} $ & $ \bm{ 0.886 \pm 0.001 } $\\
    Change & 4.9\% & & -0.1\% & & 0.1\% & \\
    \bottomrule
  \end{tabular}
  }
  \label{tab:inductive-fb}
\end{table*}

\end{document}